\documentclass[12pt,tightenlines,nofootinbib,superscriptaddress,notitlepage, APS, pra]{revtex4-1}

\usepackage{appendix}
\usepackage{amssymb}

\usepackage{amsfonts,amssymb,amstext,amsmath,amsthm,bm,slashed}
\usepackage[dvips]{graphicx}
\usepackage{mathtools}
\usepackage{color}
\usepackage{hyperref}
\usepackage{float}
\usepackage{placeins}
\usepackage{bbold}
\usepackage{indentfirst}
\usepackage{tabu}
\usepackage{verbatim}
\usepackage{comment}
\usepackage{enumerate}

\theoremstyle{definition}

\theoremstyle{plain}

\usepackage{relsize}

\newcommand{\be}{\begin{equation}}
\newcommand{\ee}{\end{equation}}
\newcommand{\barray}{\begin{array}}
\newcommand{\earray}{\end{array}}
\newcommand{\bea}{\begin{eqnarray}}
\newcommand{\eea}{\end{eqnarray}}
\newcommand{\bs}{\begin{subequations}}
\newcommand{\es}{\end{subequations}}
\newcommand{\beal}{\begin{align}}
\newcommand{\eeal}{\end{align}}

\newcommand{\DB}[1]{\lbrace #1 \rbrace_\mathrm{DB}}

\def\sig{\sigma}

\def\eps{\epsilon}

\def\lam{\lambda}
\def\Lam{\Lambda}
\def\o{\omega}

\def\a{\alpha}
\def\b{\beta}
\def\eps{\epsilon}

\def\pa{\phantom{\alpha}}
\def\bs{\bar{\sigma}}

\newcommand{\rd}{\mathrm{d}}
\newcommand{\p}{\partial}

\def\sl2c{SL(2,\mathbb{C})}
\newcommand{\vect}[1]{\left\vert #1 \right\rangle}
\newcommand{\covec}[1]{\left\langle #1 \right\vert}
\newcommand{\inner}[1]{\langle #1 \rangle}
\newcommand{\mc}[1]{\mathcal{#1}}
\newcommand{\mf}[1]{\mathfrak{#1}}

\newcommand{\abs}[1]{\vert #1 \vert}

\newcommand{\pb}[1]{\left\lbrace #1 \right\rbrace}
\newcommand*\overbar[1]{%
  \hbox{%
    \vbox{%
      \hrule height 0.5pt 
      \kern0.4ex
      \hbox{%
        \kern 0em
        \ensuremath{#1}%
        \kern 0em
      }%
    }%
  }%
} 

\newcommand{\Ad}[1] {\mathrm{Ad}_{#1}}

\def \Tr {\mathrm{Tr}}

\def \Ad{\mathrm{Ad}}

\def\bs{\bar{s}}

\newcommand{\R}{\mathbb{R}}

\usepackage{bbm}

\def\C{{\mathbbm C}}
\def\R{{\mathbbm R}}

\newcommand{\SO}{\mathrm{SO}}

\def\g{\mathfrak{g}}
\def\gg{\mathfrak{g}^*}
\def\p{\mathfrak{p}}
\def\pp{\mathfrak{p}^*}

\newcommand{\mr}[1]{\mathrm{#1}}

\usepackage{tikz}
\usetikzlibrary{decorations.pathmorphing}
\usepackage{tkz-euclide}
\usetikzlibrary{decorations.markings}
\tikzset{->-/.style={decoration={
  markings,
  mark=at position .5 with {\arrow{>}}},postaction={decorate}}}

\begin{document}
\title{Interaction Vertex for Classical Spinning Particles}

\author{{Trevor Rempel}}

\affiliation{{Perimeter Institute for Theoretical Physics, Waterloo, Ontario, Canada}}
\affiliation{\small\textit{Department of Physics, University of Waterloo, Waterloo, Ontario, Canada}}
\smallskip
\author{{Laurent Freidel}}
\smallskip 
\affiliation{{Perimeter Institute for Theoretical Physics, Waterloo, Ontario, Canada}}
\date{ \today}
\bigskip
\begin{abstract}
We consider a model of the classical spinning particle in which the coadjoint orbits of the Poincare group are parametrized by two pairs of canonically conjugate four vectors, one representing the standard position and momentum variables and the other which encodes the spinning degrees of freedom. This ``Dual Phase Space Model'' is shown to be a consistent theory of both massive and massless particles and allows for coupling to background fields such as electromagnetism. The on--shell action is derived and shown to be a sum of two terms, one associated with motion in spacetime and the other with motion in ``spin space.''  Interactions between spinning particles are studied and a necessary and sufficient condition for consistency of a three--point vertex is established. 
\end{abstract}

\maketitle

\section{Introduction}
The notion of ``intrinsic angular momentum'' was first discussed in the context of classical general relativity by Cartan \cite{cartan_1923} in 1922. Spin, as it relates to the description of elementary particles, didn't make an appearance until 1925 in the work of Goudsmit and Uhlenbeck \cite{goudsmit_1925} who proposed that the splitting of spectral lines in the anomalous Zeeman effect could be explained by attributing an internal angular momentum to the electron. This idea was made rigorous a few years later when Dirac \cite{dirac_1928} published his famous equation, now universally accepted as the correct quantum mechanical description of spin--$\frac{1}{2}$ particles. Despite the success Dirac's theory has enjoyed it offers little insight into the physical origin of spin, referred to by Pauli as a ``two--valued quantum degree of freedom.'' Modern treatments hold to this line of thought, either claiming outright that spin has no classical interpretation \cite{shankar_1994} or avoiding the topic altogether \cite{sakurai_1994}. That is not to say attempts haven't been made to understand spin from a classical perspective, the literature on the subject is vast pre--dating even Dirac.\footnote{For readers interested in the subject see the review articles \cite{nyborg_1961A, review_1996,lagrangian_1996,review_1998} or the full length books \cite{corbenbook_1968, rivas_2002}.}\\
\indent Classical models of spin can be roughly divided into two types: phenomenological and group theoretic. Phenomenological models were the first to appear and took as their starting point some intuition regarding the internal structure of a spinning particle. For example, Frenkel \cite{frenkel_1926}, Thomas \cite{thomas_1926, thomas_1927} and Kramer \cite{kramers_1934, kramers_QM} proposed that spin was represented by an anti--symmetric tensor $S_{\mu\nu}$ whose interaction with the electromagnetic field $F_{\mu\nu}$ was governed by a covariant generalization of $\partial_t{\vec{S}} \propto \vec{S}\times \vec{B}$, the equation for precession of a magnetic moment $\vec{S}$ in a magnetic field $\vec{B}$. In contrast, Mathisson \cite{mathisson_1937}, Papapetrou \cite{papapetrou_1951A, papapetrou_1951B} and Dixion \cite{dixon_1964, dixon_1967, dixon_1970} assumed that all information about the spinning particle is contained in its stress energy tensor $T_{\mu\nu}$ with equations of motion following from conservation of energy, $\nabla_\nu T^{\mu\nu}=0$. Others characterized a spinning particle by a point charge and dipole moment \cite{corben_1941,corben_1960,corben_1961}, or as a relativistic fluid \cite{weyss_1947A, weyss_1947B} while still others proposed semi--classical models \cite{barut_1958, barut_1984A}. The last of these was quantized and shown to reproduce the Dirac propagator in the path integral formalism \cite{barut_1984B, barut_1988}. This Lagrangian perspective  continues to be developped today \cite{deriglazov_2012, deriglazov_2014, costa_2012, costa_2014}.\\
\indent Group theoretic models, on the other hand, connect directly with the quantum description of a spinning particle as irreducible representations of the Poincare group. The first to attempt such a formulation were Hanson and Regge \cite{hanson_1974} and Balachandran \cite{balachandran_1976, balachandran_1979, balachandran_1981} both of whom assumed that the configuration space of a spinning particle was coordinatized by elements of the Poincare group. This approach was formalized by Kirillov \cite{kirillov_1976}, Kostant \cite{kostant_1970} and Souriau \cite{souriau_1970,souriau_1997} who showed that the coadjoint orbits of a group form a symplectic manifold and therefore have a natural interpretation as the phase space of some classical system. Several authors \cite{nielsen_1987, faddeev_1988,johnson_1989,wiegmann_1989B, mauro_2004} have utilized the coadjoint orbit method to construct classical descriptions of spin, with quantization achieved by means of the worldline formalism \cite{feynman_1950,feynman_1951}.\\
\indent This approach is dramatically different from the most common worldline treatment of spinning particles \cite{fradkin_1965, berezin_1977, howe_1976,brink_1976,schubert_2000} where the spin degrees of freedom are represented by Grassmann variables.
The group theoretical approach has in our view the merit of conceptual clarity, it allows the spinning degrees of freedom to be parametrized by variables which possess a semi--classical interpretation while also providing a common treatment of all spins at once. Moreover,  Wiegmann \cite{wiegmann_1989B} has shown the equivalence between the Grassman variable treatment and the bosonic group theoretical approach.
In this work we focus on the worldline description of spinning particles in terms of realistic compact degrees of freedom. \\
\indent While providing some deep insight into the origin of spin, to the authors knowledge, the worldline approach has been principally concerned with understanding the propagation and quantization of single particles, possibly coupled to background fields. The question of interactions between several worldlines has not yet been developed and will be a focus of the present paper.  One motivation for exploring this topic came from the recent understanding that 
 there is, for a spinless particle, a direct connection between the form of the interaction vertex in momentum space and the locality of the corresponding interaction. Specifically, following \cite{relative_locality}, we showed in \cite{action_vertices} that a sufficient condition for local interactions is that the vertex factor be linear in the constituent momenta. Conversely, theories such as relative locality \cite{relative_locality, deepening, scalar_curved} which permit momentum combination rules that are non--linear also realize non--local interactions. It is well known that for a scalar particle the vertex factor is just conservation of momenta and interactions are local as expected. Spinning particles are a different story since the presence of internal angular momentum modifies the vertex factor and the localisation properties \cite{Freidel:2007qk}. The question which we wish to answer is whether this modification is dramatic enough to allow for non--local behaviour.  \\
\indent The first result of this paper is to propose a new parametrisation for the phase space of a relativistic spinning particle, called the Dual Phase Space Model (DPS). In this parametrization the standard phase space of $(x,p)$ is extended by a second set of canonical variables $(\chi, \pi)$ which span a ``spin'' or ``dual'' phase space. We describe in detail the set of constraints on this dual space that realizes the relativistic spinning particle and show that interactions are local not only in $x$ but in the dual position space $\chi$ as well.
This dual locality property is one of the main results of this paper.
We also provide a precise formulation of the on--shell action for a spinning particle.
From this analysis it becomes clear that the quantum phase factor is the sum of two terms: one is the usual proper--time $\tau=m\int |\dot{x}| \rd t$, which depends on the spacetime motion,  while the second factor is an angle $\theta=s\int |\dot{\chi}|\rd t$ that depends on the spin motion.\\
\indent This paper is organized as follows: In Section \ref{sec:elementary} we present an overview of the coadjoint orbit method for a general matrix Lie group, the procedure is then implemented explicitly in Section \ref{sec:coadjoint} for the Poincare group. Two possible parametrizations of the these orbits are presented in Section \ref{sec:models} and the spin quantization condition is discussed. Section \ref{sec:DP} introduces the Dual Phase Space Model which is shown to be a consistent description of both massive and massless spinning particles. Equations of motion are derived, the on--shell action is calculated and a consistent coupling to electromagnetism is demonstrated. Finally, interactions between spinning particles are studied in Section \ref{sec:vertex}, where we prove that dual locality is a necessary and sufficient condition for consistency of the three--point vertex.

\section{Elementary Classical Systems and Their Quantization}\label{sec:elementary}
In this section we discuss the mathematical preliminaries which allow for a classical formulation of the spinning particle. For some readers this might sound paradoxical since spin is often viewed as a purely quantum object. However, while there are some phenomena, like the relationship between spin and statistics, which are purely quantum, it does not follow that the relativistic spinning particle has no classical description. What it does mean is that this description will only be accurate in the limit of large spins.\\
\indent  It is generally true that one can construct a classical realization of any quantum structure associated with a group $G$; for spin the relevant group is  the Poincar\'e group. The procedure for doing so is called the coadjoint orbit method \cite{kirillov_2002} and is outlined below for the case of matrix Lie groups; a reasonable simplification as most groups of interest fall into this category.\\
 \indent Let $G \subset \mathrm{GL}(n,\C)$ be a matrix Lie group and $\g \subset \mathrm{Mat}(n,\C)$ its Lie algebra, the adjoint action of $g \in G$ on $X \in \g$ is then matrix conjugation $\Ad(g)X = gXg^{-1}$ and the coadjoint action of $G$ on the dual algebra $\gg$ is obtained by taking the dual of $\Ad$. It satisfies
\begin{align}\label{coadjoint}
\inner{\Ad^*(g)\lam, X} = \inner{\lam, \Ad(g^{-1})X},
\end{align}
where $\lam \in \gg$ and $\inner{,}$ denotes the natural pairing between $\g$ and $\gg$. Each coadjoint orbit $\mc{O}_\lam = \lbrace \Ad^*(g)\lam \, \vert \, g \in G\rbrace$ possesses a natural symplectic structure $\sig_\lam$ and the pair $(\mc{O}_\lam, \sig_\lam)$ forms the classical phase space associated with the symmetry group $G$. To obtain $\sig_\lam$ explicitly we let $H_\lam$ be the isotropy group for some $\lam \in \gg$, then the bijection $p_\lam: G/H_\lam \to \mc{O}_\lam : [g] \to \Ad^*(g)\lam$ identifies the homogeneous space $G/H_\lam$ with the coadjoint orbit through $\lam$. A choice of section $g: G/H_\lam \to G$ allows us to pull back the Maurer--Cartan form on $G$ to give a symplectic potential on $G/H_\lam$
\begin{align}\label{thetasym}
\theta_\lam = \inner{\lam, g^{-1}dg}.
\end{align}
The value of $\theta_\lam$ depends explicitly on the choice of section. In particular, if $h: G/H_\lam \to H_\lam$ the change of section $g \to g h$ yields a corresponding variation $\delta \theta_\lam =-\inner{\lam , h^{-1}dh}$. Since $Ad^*(H_\lam) \lambda=\lam$, this sectional dependence disappears when considering the symplectic form
\begin{align}\label{o-symplectic}
\o_\lam = d\theta_\lam = -\inner{\lam, g^{-1}dg \wedge g^{-1} dg},
\end{align}
where the Maurer--Cartan equation $d(g^{-1}dg) =- g^{-1}dg \wedge g^{-1} dg$ has been used. One can now obtain the symplectic form on $\mc{O}_\lam$ by taking the pull back of $\o_\lam$ under $p_\lam^{-1}$: $\sig_\lam = (p_\lam^{-1})^*\o_\lam$. \\
\indent We can proceed a bit further. Let $\hat{X}$ denote the extension of the Lie algebra element $X \in \g$ to a right invariant vector field over $G$, then
\begin{align}
\o_\lam(\hat{X},\cdot) = \inner{F_\lambda(g), [X,dg g^{-1}]} = d\inner{F_\lambda,X},
\end{align}
where $F_\lambda(g) := \Ad^*(g)\lam$ is a generic element of the coadjoint orbit through $\lam$. It follows that the linear function $\mc{H}_X(g) := \inner{F_\lambda(g),X}$ is a Hamiltonian for the group action and 
$F_\lambda: G/H_\lambda \to \gg$ is its moment map. It follows that the Poisson bracket between two such functions is the commutator $\pb{\mc{H}_X,\mc{H}_Y} = \mc{H}_{[X,Y]}$.\\
\indent A classical description of some system is only useful if one can pass to the corresponding quantum version. In the present context this transition amounts to finding a map between the coadjoint orbits of a group and its irreducible representations. The key idea is that a classical phase space corresponds to a quantum Hilbert space  and a phase space function to an operator, the symmetry then restricts the mapping almost uniquely. A formal correspondence between a classical system and its quantum counter--part is accomplished via geometric quantization \cite{woodhouse_1991} which also forms the basis of the Feynman path integral formulation of quantum mechanics. If the quantum system is finite dimensional the corresponding phase space has to be compact since the Hilbert space dimension is related to the phase space volume. Heuristically, the construction proceeds as follows: Let $\mc{O}_\lam$ be a coadjoint orbit of $G$ and let $X \in \g$ be a Lie algebra element, the trace of a group element in a unitary irreducible representation $\rho_\lam: G \to \mc{O}_\lam$ of highest weight $\lam$ is then given by
\begin{align}
\Tr_{V}\left(\rho_\lam(e^{iX})\right) = \int \mc{D}g\, e^{\frac{i}{\hbar}\int_{S^1}\left[\inner{\lam, g^{-1}\dot{g}} - \inner{F_\lam(g),X}\right]d\tau},
\end{align}
where the path integral is taken over all group valued periodic maps $g: S^1 \to G$. This is just a generalization of the usual Feynman path integral quantization where $\Tr e^{i\hat{\mc{H}}(\hat{p},\hat{q})}$ is written as
\begin{align}
\Tr e^{i\hat{\mc{H}}(\hat{p},\hat{q})} = \int \mc{D}p \mc{D}q e^{\frac{i}{\hbar}\int_{S^1} (p\dot{q} - H(p,q))d\tau},
\end{align}
and the paths are chosen to be periodic. Here the phase space variables are $(p,q)$, with symplectic potential $p \rd q$ and Hamiltonian $\mc{H}(p,q)$. In our case, the phase space variables are group elements $g$, with symplectic potential $\theta_\lam = \inner{\lam, g^{-1}\rd g}$ and Hamiltonian $\mc{H}_X(g)= \inner{F_\lambda(g), X}$ as discussed above.\\
\indent This procedure can be reversed, mapping irreducible representations onto coadjoint orbits. To see this, suppose that $\rho: G \to \mathrm{GL}(V)$ is a unitary irreducible representation of $G$ over the vector space $V$. To each normalised vector $\vect{\Lam} \in V$,  we can associate a linear functional $\lam \in \gg$ by defining
\begin{align}
\lam(X) := \hbar \covec{\Lam}\rd \rho(X)\vect{\Lam},
\end{align}
where $X \in \g$ and $\rd \rho$ is the representation of $\g$ induced by $\rho$.
$H_\lam$ is by definition the subgroup that acts diagonally on $\Lambda$ and so, if $h = e^{iH/\hbar} \in H_\lam$ its action is given by 
\be
\rho(h) \vect{\Lam} =  e^{i\frac{\lam(H)}{\hbar}} \vect{\Lam}.
\ee
 It follows that the linear functional associated with $\rho(g)\vect{\Lam}$ is $\Ad^*(g)\lam$. If $\rho$ is an irreducible representation, every vector in $V$ 
can be represented as a linear combination of elements $\rho(g)\vect{\Lam}$ with $g \in G$, therefore  the map 
\bea V &\to & \mc{O}_\lam, \\
\rho(g)\vect{\Lam} &\mapsto& F_\lambda(g),
\eea identifies rays in  $V$ with points in the coadjoint orbits. More explicitly, if we label elements of $\mc{O}_\lam$ by the operators ${X}_\rho(g) := \rho(g)\vect{\Lam}\covec{\Lam}\rho^\dagger(g)$, then the symplectic form 
\be\label{qsymplectic}
{\omega}_\rho := -\hbar\Tr_V({X}_\rho \rd {X}_\rho\wedge \rd {X}_\rho),
\ee simplifies to
$
{\omega}_\rho 
=\hbar \covec{\Lam}\rho(g^{-1})d\rho(g) \wedge \rho(g^{-1})d\rho(g) \vect{\Lam}
$
which is equivalent to the one given in \eqref{o-symplectic}.

\section{Coadjoint Orbits of the Poincare Group}\label{sec:coadjoint}

Although we have presented the coadjoint orbit method in general we are only interested in its application to the Poincare group $\mc{P} = \SO(3,1) \rtimes \R^4$, which is well known to describe the symmetries of a relativistic spinning particle.  In this section we will review the construction of these orbits and show that they are characterized by two quantities which are identified with the particle's mass and spin. \\
\indent Let $g(\Lambda, x)$ be a generic element of the Poincare group, where $\Lambda \in \SO(3,1)$ is a Lorentz transformation and $x \in \R^4$ a translation, the group product is given by $(\Lambda_1,x_1)(\Lambda_2,x_2) = (\Lambda_1\Lambda_2, x_1 + \Lambda_1x_2)$. The generators of translations and Lorentz transformations, which form a basis for the Lie algebra $\mf{p}$, are denoted $P_\mu$ and $\mc{J}_{\mu\nu} = -\mc{J}_{\nu\mu}$ respectively and satisfy
\begin{align*}
[P_\mu, \mc{J}_{\nu\rho}] = \eta_{\mu\nu}P_\rho - \eta_{\mu\rho}P_\nu, \qquad [\mc{J}_{\mu\nu},\mc{J}_{\rho\sig}] =  \eta_{\mu\sig}\mc{J}_{\nu\rho} + \eta_{\nu\rho}\mc{J}_{\mu\sig} - \nu_{\mu\rho}\mc{J}_{\nu\sig} - \eta_{\nu\sig}\mc{J}_{\mu\rho} .
\end{align*} 
It is now a straightforward exercise to compute the adjoint action of $g(\Lambda, x)$ on $\mf{p}$, viz
\begin{align}\label{PonAdjoint}
\Ad(g(\Lambda,x))P_\mu = \Lambda^\nu_{\pa \mu}P_\nu, \qquad \Ad(g(\Lambda, x))\mc{J}_{\mu\nu} = \Lambda_{\pa \mu}^{\rho} \Lambda_{\pa \nu}^{\sig}\left(\mc{J}_{\rho\sig} + P_\rho x_\sig - P_\sig x_\rho\right),
\end{align}
Introduce dual generators $\hat{P}^\mu $ and $\hat{\mc{J}}^{\mu\nu}$ as a basis for the dual algebra, $\pp$, and let $\inner{,}$ be the natural pairing between $\p$ and $\pp$, then $\inner{\hat{P}^\mu ,P_\nu} = \delta^\mu _\nu$ and $\inner{\hat{\mc{J}}^{\mu\nu},\mc{J}_{\rho\sig}} = 2\delta^\mu _{[\rho}\delta^\nu_{\sig]}$. The coadjoint action is obtained from equation \eqref{PonAdjoint} by recalling its definition in terms of the adjoint action, see \eqref{coadjoint}. We find
\begin{align}
\Ad^*(g(\Lambda,x))\hat{P}^\mu  = \Lambda^{\pa \mu}_{\nu}\left(\hat{P}^\nu - x_\rho\hat{\mc{J}}^{\nu\rho}\right), \qquad \Ad^*(g(\Lambda,x))\hat{\mc{J}}^{\mu\nu} = \Lambda^{\pa \mu}_{\rho}\Lambda^{\pa \nu}_{\sig}\hat{\mc{J}}^{\rho\sig},
\end{align}
where $\Lambda_\mu^{\pa \nu} = (\Lambda^{-1})^\nu_{\pa \mu}$.
Elements of the dual algebra $F \in \pp$ are parametrized by a vector $m_\mu$ and an anti--symmetric tensor $M_{\mu\nu}$, $F = (m_\mu,M_{\mu\nu})$. Under the coadjoint action these components transform as
\begin{align}
m	_\mu &\xrightarrow{\Ad^*(g(\Lambda,x))}  p_\mu = \Lambda^{\pa \nu}_{\mu}m_\nu,\label{pcoadjoint}\\
M_{\mu\nu} &\xrightarrow{\Ad^*(g(\Lambda,x))} J_{\mu\nu} = (x\wedge p)_{\mu\nu} + \Lambda^{\pa \rho}_{\mu}\Lambda^{\pa \sig}_{\nu}M_{\rho\sig}\label{mcoadjoint},
\end{align}
where $(A\wedge B)_{\mu\nu} = A_\mu B_\nu - A_\nu B_\mu$. The quantities $p_\mu$ and $J_{\mu\nu}$ have standard physical interpretations, $p_\mu$ represents the total linear momentum of the particle while $J_{\mu\nu}$ represents the total angular momentum about the origin. Notice that we can split the total angular momentum as $J = L + S$, where $L = x\wedge p$ is the orbital part and $S =(\Lam  M \Lam^T)$ is the spin angular momenta.\\ 
\indent These orbits are characterized by the value of two invariants\footnote{Quantities which remain unchanged by the coadjoint action of $\mc{P}$.}, one of which is $p^2 = -m^2$, with $m$ representing the mass of the particle. If $m>0$, the other invariant is $w^2 =  m^2s^2$ where $w_\mu = \frac{1}{2 \hbar }\eps_{\mu\nu\rho\sig}p^\nu J^{\rho\sig}$ is the Pauli--Lubanski vector and $s$ is identified with the particles spin. The phase space for a relativistic spinning particle of mass $m$ and spin $s$ is then
\begin{align}\label{constraints}
\mc{O}_{m,s} = \lbrace (p_\mu,J_{\mu\nu}) \,\vert\, p^2 = -m^2 \,\mr{and} \, w^2 = m^2s^2\rbrace.
\end{align}
An arbitrary element $F_{m,s} \in \mc{O}_{m,s}$ defines the symplectic form $\sig_{F_{m,s}}$ and the symplectic manifold $(\mc{O}_{m,s} , \sig_{F_{m,s}})$ constitutes a complete description of the relativistic spinning particle. \\
\indent If, on the other hand, $m = 0$ then $w^2 = 0$ and since $w\cdot p = 0$ the Pauli--Lubanski vector must be proportional to the momentum $w_\mu = s p_\mu$; the constant of proportionality will be the second orbit invariant. Physically, this represents a massless spinning particle with helicity given by $s$; the corresponding phase space is denoted $(\mc{O}_{0,s},\sig_{F_{0,s}})$. There should be no confusion in denoting the spin and helicity by the same variable $s$ as it will be clear from context what is being referred to. 

\section{Models of the Classical Spinning Particle}\label{sec:models}
Given a coadjoint orbit of the Poincare group, equation \eqref{constraints}, a model of the relativistic spinning particle is obtained by making a choice of coordinates on $\mc{O}_{m,s}$. There are many viable options and the resulting theories can seem disparate, but this is only superficial as one can always find a coordinate transformation between competing models. We demonstrate this explicitly for two popular coordinatizations, those of Balachandran \cite{balachandran_1979} and Wiegmann \cite{wiegmann_1989B} and in the process examine how the quantization condition $2s \in \mathbb{Z}$ arises.
\subsection{Homogeneous Space}\label{sec:homogeneous}
With $m>0$ we can choose $F_{m,s}$ to have components $m_\mu:= m\delta_\mu^0$ and $M_{\mu\nu}:= 2\hbar s\delta^1_{[\mu}\delta^2_{\nu]}$ which transform under the coadjoint action of $g(\Lambda, x)$ as
\begin{align}
m\delta_\mu^0 &\longrightarrow p_\mu= m\Lambda^{\pa 0}_{\mu},\\
2s\delta^1_{[\mu}\delta^2_{\nu]} &\longrightarrow J_{\mu\nu}= 2m x_{[\mu}\Lambda^{\pa 0}_{\nu]} +2\hbar s\Lambda_{[\mu}^{\pa 1}\Lambda_{\nu]}^{\pa 2}.
\end{align}
The phase space $\mc{O}_{m,s}$ is then regarded as a subset of $\mc{P}$ coordinatized by $\pb{x^\mu ,\Lambda_{\mu}^{\pa 0},\Lambda_{\mu}^{\pa 1},\Lambda_{\mu}^{ \pa 2}}$. In this parametrization the splitting $J = L + S$ is realized explicitly as  
\begin{align}\label{spinP}
L_{\mu\nu} = m\left(x_\mu\Lambda^{\pa 0}_{\nu} - x_\nu \Lambda^{\pa 0}_{\mu}\right) \qquad \mr{and} \qquad S_{\mu\nu} = \hbar s\left(\Lambda_{ \mu}^{\pa 1}\Lambda_{\nu}^{\pa 2} - \Lambda_{\nu}^{\pa 1}\Lambda_{\mu}^{\pa 2}\right).
\end{align} 
Comparison with equations (1) and (2) of \cite{balachandran_1979} shows that this parametrization is identical to that of Balachandran. \\
\indent  To obtain the symplectic potential $\theta_{m,s}$ we first expand the Lie algebra valued one form $g^{-1}dg$ in the basis $\lbrace P_\mu,\mc{J}_{\mu\nu}\rbrace$
\begin{align*}
g^{-1}(\Lambda, x)dg(\Lambda, x) = -\Lambda_\nu^{\pa \mu}dx^\nu P_\mu + \frac{1}{2}\eta_{\rho\sig}\Lambda^{\rho\mu}d\Lambda^{\sig\nu}\mc{J}_{\mu\nu},
\end{align*} 
then with $F_{m,s}$ as described above, equation \eqref{thetasym} gives
\begin{align}\label{Ppotential}
\theta_{m,s} = -m\Lambda_\mu^{\pa 0} dx^\mu  +  \frac{\hbar s}{2}\eta^{\mu\nu}\left(\Lambda_\mu^{\pa 1}d\Lambda_\nu^{\pa 2} - \Lambda_\nu^{\pa 2}d\Lambda_\mu^{\pa 1}\right).
\end{align}
We can now identify $p_\mu = m\Lambda_\mu^{\pa 0}$ with the momentum conjugate to $x^\mu $ and write the symplectic form $\o_{m,s} = d\theta_{m,s}$ as 
\begin{align}\label{Pform}
\o_{m,s} = dx^\mu \wedge dp_\mu + \hbar s\eta^{\mu\nu}d\Lambda_\mu^{\pa 1}\wedge d\Lambda_\nu^{\pa 2}.
\end{align}
Finally, we obtain an action by regarding all coordinates as a function of an auxiliary parameter $\tau$ and integrate the symplectic potential, viz
\begin{align}\label{Paction}
S = \int d\tau\left[p_\mu\dot{x}^\mu  - \frac{\hbar s}{2}\eta^{\mu\nu}\left(\Lambda_\mu^{\pa 1}\dot{\Lambda}_\nu^{\pa 2} - \Lambda_\nu^{\pa 2}\dot{\Lambda}_\mu^{\pa 1}\right)\right],
\end{align} 
where we have dropped an overall minus sign in the action. Note that we still regard $p_\mu$ as being derived from the Lorentz transformation $\Lambda_\mu^{\pa 0}$ which implies that this parametrization is explicitly on--shell.

\subsection{Vector on a Sphere}\label{sec:vectsphere}
In \cite{wiegmann_1989B} Wiegmann parametrizes, in a natural way, the spinning degrees of freedom by a unit vector $n_\mu$ orthogonal to the linear momentum $p_\mu$. 
We now explicitly show that the Wiegmann parametrisation is equivalent to Balachandran. To see how this correspondence comes about, put $A_\mu = \Lambda_\mu^{\pa 1}$ and $B_\mu = \Lambda_\mu^{\pa 2}$, then
\begin{align}\label{sigS}
\o^\mr{S}_{m,s} &= \hbar s\eta^{\mu\nu}dA_\mu \wedge dB_\nu, \qquad \mr{and} \qquad S_{\mu\nu} = \hbar s(A\wedge B)_{\mu\nu},
\end{align}
where $\o_{m,s}^\mr{S} =\o_{m,s}  - dx\wedge dp$ is the spin component of the symplectic potential. Introduce the unit momenta $u_\mu = p_\mu/m$ and define $n_\mu = \eps_{\mu\nu\rho\sig}u^\nu A^\rho B^\sig$; note that $n_\mu$ is proportional to the Pauli--Lubanski vector $w_\mu = ms n_\mu$. The set $\lbrace u_\mu,n_\mu,A_\mu,B_\mu\rbrace$ forms an orthonormal basis for $\R^4$ adapted to the particle's motion. We can then expand the Minkowski metric as
\begin{align*}
\eta_{\mu\nu} = -u_\mu u_\nu + n_\mu n_\nu + A_\mu A_\nu + B_\mu B_\nu.
\end{align*}
If one replace the $\eta^{\mu\nu}$ appearing in $\o_{m,s}^\mr{S}$ with the expanded version above we obtain
\begin{align*}
\o_{m,s}^\mr{S} &= \frac{\hbar s}{2}(A\wedge B)^{\mu\nu}\left(du_\mu\wedge du_\nu - dn_\mu\wedge dn_\nu\right).
\end{align*}
We can now make use of the relation $(A\wedge B)_{\mu\nu} = -\eps_{\mu\nu\rho\sig}u^\rho n^\sig$ to eliminate $A$ and $B$ from the expressions for $\o_{m,s}^\mr{S}$ and $S_{\mu\nu}$ and obtain a parametrization given entirely in terms of $u_\mu$ and $n_\mu$:
\be
\o^\mr{S}_{m,s} = \frac{\hbar s}{2}\eps_{\mu\nu\rho\sig}u^\mu n^\nu \left(dn^\rho\wedge dn^\sig - du^\rho\wedge du^\sig\right)\label{nsymplectic}, 
\qquad
S^\mr{S}_{\mu\nu} = -\hbar s\eps_{\mu\nu\rho\sig}u^\rho n^\sig,
\ee
which corresponds to the Wiegmann expressions  \cite{wiegmann_1989B}. The phase space of this model is coordinatized by $\pb{x^\mu ,p^\mu ,n^\mu }$ subject to the constraints
\begin{align}
p^2 = -m^2, \qquad n^2 = 1, \qquad p\cdot n = 0,
\end{align}
which define the on--shell hypersurface. In the rest frame $u_\mu = \delta^0_\mu$ and the symplectic form $\o_{m,s}^\mr{S}$ reduces to
\begin{align}\label{s2form}
\sig^\mr{S} = -\frac{\hbar s}{2}\eps_{ijk}n^idn^j\wedge dn^k,
\end{align}
which is just the area form on a sphere of radius $\hbar s$. It follows that we can regard the two--form \eqref{nsymplectic} as a ``relativistic generalization'' of the symplectic structure on a sphere and $n_\mu$ as an $S^2$ vector boosted in the direction of $p_\mu$. 

\subsection{Quantization Condition}
As presented above, the quantity $s$, which represents the particles spin, is permitted to assume any real value. To recover the usual restriction -- $2s \in \mathbb{N}$ -- one demands that the symplectic form $\o/ \hbar$ is integral, i.e. the integral of $\o/\hbar$ over a non--trivial two cycle is an integer multiple of $2 \pi$. Consider what this means for the model of Section \ref{sec:vectsphere} where there is a single non--trivial two cycle, namely the sphere $S^2$. In the rest frame, the quantization condition says
\begin{align*}
\frac1{\hbar} \int_{S^2} \o = s\int_{S^2} \frac{1}{2}\eps_{ijk}n^idn^j\wedge dn^k \in 2\pi \mathbb{N}.
\end{align*}
The quantity under the integral sign is the area form on the two--sphere and evaluates to $4\pi$ which immediately gives the expected result $2s \in  \mathbb{N}$.\\
\indent A more intuitive approach is as follows: Let $\mc{C}$ denote the worldline of a spinning particle, then one could attempt to define an action as the integral over the symplectic potential, i.e. $S = \int_\mc{C} \theta_{m,s}$. Unfortunately, this is not well defined since the symplectic form is not exact and so $\theta_{m,s}$ does not exist globally. Instead we need to define $S$ as the integral of $\o_{m,s}$ over some surface of which $\mc{C}$ is a boundary
\begin{align*}
S = \int_\mc{C} \theta_{m,s} = \int_\mc{S} \o_{m,s},
\end{align*}
where $\partial \mc{S} = \mc{C}$. The choice of $\mc{S}$ is ambiguous 
but if we demand that different surfaces change $S$ by a multiple of $2\pi \hbar$ then the path integral will be well defined, since it is $e^{\frac{i}{\hbar} S}$ which is the relevant quantity. For the vector on a sphere,  $\mc{C} = S^1$ and so $\mc{S}$ can either be the upper or lower half sphere. In the rest frame we have 
\begin{align*}
\int_{S^2_{\mr{upper}}} \o_{m,s}^\mr{S} = \int_{S^2} \o_{m,s}^\mr{S} + \int_{S^2_\mr{lower}} \o_{m,s}^\mr{S},
\end{align*}
and so we demand that $\int_{S^2} \o_{m,s}^\mr{S} = 2\pi \hbar $, which is the same condition arrived at in the more formal approach.

\section{Dual Phase Space Model}\label{sec:DP}

The previous section presented a sampling of possible parametrizations for the coadjoint orbits of the Poincare group. There are many other options, all of which are equivalent and can be used interchangeably depending on what aspect of the theory is to be emphasised. Presently, our interest is in analysing the interaction vertex and so we introduce a parametrization that is particularly well suited to this task.

\subsection{Choosing the Coordinates}
To define this parametrization we introduce a length scale $\lambda$ and an energy scale $\epsilon$ such that $ \lambda \epsilon = \hbar$, otherwise these scales are arbitrary constants.
Recall the parametrization presented in \ref{sec:homogeneous} and define variables $\chi_\mu = \lambda  \Lambda_\mu^{\pa 1}$ and $\pi_\mu = \epsilon {s}\Lambda_\mu^{\pa 2}$ so that the symplectic form \eqref{Pform} is written as\footnote{From now on we will drop subscripts on the symplectic form.}
\begin{align}\label{dpsymplectic}
\o &=dx^\mu \wedge dp_\mu + d\chi^\mu  \wedge d\pi_\mu.
\end{align}
We now forget that $p_\mu$, $\chi^\mu $ and $\pi_\mu$ are components of a Lorentz transformation and instead regard them as variables on a classical phase space coordinatized by $\pb{x^\mu ,p_\mu, \chi^\mu ,\pi_\mu}$. It follows from \eqref{dpsymplectic} that $(x^\mu , p_\mu)$ and $(\chi^\mu , \pi_\mu)$ form pairs of canonically conjugate variables with Poisson brackets
\begin{align}\label{pb}
\pb{x^\mu ,p_\nu} = \delta^\mu _\nu,  \qquad \pb{\chi^\mu ,\pi_\nu} = \delta^\mu _\nu,
\end{align}
all others vanishing. From this perspective $\chi^\mu $ and $\pi_\mu$ span a ``dual'' phase space, separate from the standard phase space of $x^\mu $ and $p_\mu$, which encodes information about the particles spin. The internal angular momentum, $S_{\mu\nu}$, further bears out this duality since in these variables it assumes the form (see \eqref{spinP})
\begin{align}\label{DSPspin}
S_{\mu\nu} = (\chi\wedge \pi)_{\mu\nu},
\end{align}
in direct analogy to orbital angular momentum $L_{\mu\nu} = (x\wedge p)_{\mu\nu}$. It is for this reason that we have called this formulation the \textit{Dual Phase Space Model} or DPS and view $\chi_\mu$ and $\pi_\mu$ as a dual ``coordinate'' and ``momenta'' respectively. \\
\indent It remains to explicitly impose relations among the phase space variables that were implicit in their origin as Lorentz transformations. These constraints will define the dynamics of our theory and are given by
\begin{align}\label{DPconstraints}
\left(p^2 = -m^2,\;\;\pi^2 = \epsilon^2 s^2 \right), \qquad   \left( p \cdot \pi^{} = 0,\;\; p \cdot \chi = 0\right) , \qquad \left( \chi^2 = \lambda^2, \;\; \chi\cdot \pi = 0\right).
\end{align}
We have grouped the constraints in this manner to emphasise the duality mentioned above. The first pair are mass shell conditions, one in standard phase space $p^2 = -m^2$ and one in dual phase space $\pi^2 = \epsilon^2s^2$. In this description the spin is proportional to the length of the dual momenta. In the second set we see that the two phase spaces are not independent, rather dual phase space is orthogonal to the canonical momenta. The final two constraints emphasise the dramatic difference between standard phase space and dual phase space, since in the former $x$ is totally unconstrained, while $\chi$ is constrained to live on a $2$-sphere. \\
\indent As presently formulated DPS assumes $m \neq 0$; recall that we made this assumption at the outset of section \ref{sec:homogeneous}. This restriction can easily be lifted as all aspects of the current formulation, both Poisson brackets and constraints, are well defined in the limit $m \to 0$. 

An important point to emphasize is that this parametrization is invariant under an SL$(2,\mathbb{R})$ global symmetry, since any transformation of the form
\be 
( {\chi}_\mu,  {\pi}_\mu) \to (A  \chi_\mu + B  \pi_\mu ,C \chi_\mu + D  \pi_\mu ),\qquad AD-BC=1,
\ee
 does not alter the Poisson brackets (\ref{pb}) or  the angular momenta (\ref{DSPspin}). Part of this symmetry can be fixed by imposing the  orthogonality condition $\pi\cdot \chi =0$, the remaining symmetry consists of a re--scaling $( {\chi}_\mu,  {\pi}_\mu) \to (\alpha  \chi , \alpha^{-1} {\pi})$ as well as a rotation
\be 
( {\chi}_\mu,  {\pi}_\mu) \to (\cos \theta  \chi_\mu + \tfrac{\lambda}{\epsilon s}\sin\theta   \pi_\mu, 
 \cos\theta  \pi_\mu- \tfrac{\epsilon s}{\lambda}\sin \theta  \chi_\mu ).
\ee
These demonstrate, respectively, that the choice of scales $\lam$ and $\eps$ as well as the initial direction of the dual momenta are immaterial, only the product $\lam\eps$ is physically meaningful. We now assume that a choice of scale and axis has been made.\\
\indent A brief note before we continue: The parametrization presented in this section is identical to the one used by Wigner in his description of continuous spin particles \cite{wigner_1948}, see also \cite{edgren_2005} for a classical realization which emphasis the similarity. However, to the authors knowledge it has never been used in the context of standard spinning particles.

\subsection{Action and Equations of Motion}
An action for DPS is obtained by making the appropriate change of variables to equation \eqref{Paction}, and explicitly implementing the constraints \eqref{DPconstraints} by means of Lagrange multipliers, viz
\begin{align}\label{dps-action}
S &= \int d\tau \left[p_\mu\dot{x}^\mu  + \pi_\mu\dot{\chi}^\mu  - 
\frac{N}{2}(p^2 +m^2) - 
\frac{M}{2}\left( \frac{\pi^2}{\epsilon^2} + \frac{s^2\chi^2}{\lambda^2} -2s^2\right)\right.\\
& \qquad \quad- \left. \frac{N_1}{2}\left(\frac{s^2\chi^2}{\lambda^2} - \frac{\pi^2}{\epsilon^2}  \right) 
- {N_2}\left({\chi\cdot\pi}\right) - N_3\left({p\cdot \pi}\right) - N_4\left({p\cdot \chi}\right)  \right]\nonumber,
\end{align}
where we have combined some of the constraints in anticipation of the upcoming constraint analysis. Computing the constraint algebra we find, for $ms \neq 0$, there are {\it two first class constraints}
\begin{align}
\Phi_{\mr{m}} : = \frac{1}{2} \left(p^2 + m^2 \right),\qquad
\Phi_{\mr{s}} : = \frac{1}{2} \left( \frac{\pi^2}{\epsilon^2} + \frac{s^2\chi^2}{\lambda^2} \right) - s^2 ,
\end{align}
and {\it four second class constraints}
\begin{align}
\Phi_1&= \frac{1}{2}\left( \frac{s^2\chi^2}{\lambda^2}  - \frac{\pi^2}{\epsilon^2} \right), &  \Phi_2 &= \chi\cdot\pi,\\
\Phi_3 &=p\cdot \pi, & \Phi_4 &= p\cdot \chi.
\end{align} 
The latter satisfy the algebra
\begin{gather*}
\pb{\Phi_{1}, \Phi_{2} } \approx 2s^2 , 
\qquad\pb{\Phi_{3}, \Phi_{4}} \approx m^2,
\end{gather*}
where $\approx$ denotes equality on the constraint surface and all other commutators vanish\footnote{The off-shell algebra is a semi-direct product of $\mr{SL}(2,\mathbb{R})$ with the 2-dimensional Heisenberg algebra $H_2$.  The $\mr{SL}(2,\mathbb{R})$ algebra
 consists of  $\hbar(\Phi_{\rm{s}}+s^2)$, $\hbar\Phi_{1}$ and $\Phi_2$.
\bea
\pb{\Phi_{1}, \Phi_{2} } = 2(\Phi_{\rm{s}}+s^2),\quad  
\pb{\Phi_{\rm{s}}, \Phi_{1} } &=& -2\Phi_2/\hbar^2,\quad  
\pb{\Phi_{\rm{s}}, \Phi_{2} } = 2\Phi_1.
\eea
These in turn act naturally on $\Phi_3$ and $\Phi_4$ 
\bea
\pb{\Phi_{\rm{s} }, \Phi_{3} } = \frac{\Phi_{4}}{\epsilon^2} ,\quad
\pb{\Phi_{1}, \Phi_{3} } &=&  \frac{\Phi_{4}}{\epsilon^2} ,\quad 
\pb{\Phi_{2}, \Phi_{3} } = \Phi_{3}.\\
\pb{\Phi_{\rm{s} }, \Phi_{4} } = -\frac{s^2}{\lambda ^2}\Phi_{3} ,\quad
\pb{\Phi_{1}, \Phi_{4} } &=&  \frac{\Phi_{3}}{\lambda^2} ,\quad 
\pb{\Phi_{2}, \Phi_{4} } = -\Phi_{4}.
\eea
while together $\Phi_3$ and $\Phi_4$ satisfy
\bea
\pb{\Phi_{3}, \Phi_{4} } =  (m^2- 2\Phi_{\rm{m}}) .
\eea
}. This means that $(\Phi_1,\Phi_2)$ form a canonical pair whenever $s\neq 0$, as do $(\Phi_3,\Phi_4)$ when $m\neq 0$. Furthermore, when $m = 0$ the constraints $\Phi_3$ and $\Phi_4$ become first class and so a massless spinning particle is described by {\it four} first class constraints and {\it two} second class constraints. For completeness we have included an explicit expression for the Dirac brackets in Appendix \ref{app:dirac-brackets}.
\\
\indent The momentum constraint $\Phi_{\mr{m}}$ generates, as usual, the re--parametrisation invariance of the worldline $\delta  {x}_\mu= -N  {p}_\mu$. On the other hand, the spin constraint $\Phi_{\mr{s}}$ generates a $U(1)$ gauge transformation of the $\chi$ and $\pi$ variables. This transformation rotates the dual variables while preserving their normalization constraints $\Phi_i$:
 \be\label{gauge}
 \delta  {\pi}_\mu = +\left(\frac{s^2M}{\lambda^2}\right)  \chi_\mu,\qquad
 \delta  {\chi}_\mu = -\left(\frac{M}{\epsilon^2}\right)  \pi_\mu.
 \ee 

\subsection*{Massive Spinning Particle}
Let's now assume that $m \neq 0$, then the constraints $\Phi_i$, $i=1,\cdots, 4$ are second class and so the associated Lagrange multipliers, $N_1,N_2,N_3,N_4$ must vanish. The resulting Hamiltonian is given by
\be
H = {N}\Phi_{\rm{m}} + {M}\Phi_{\mr{s}}= 
\frac{N}{2} \left(p^2 + m^2 \right)+ \frac{M}{2} \left( \frac{\pi^2}{\epsilon^2} + \frac{s^2\chi^2}{\lambda^2}  - 2s^2\right).
\ee
and defines time evolution in the standard fashion: $\dot{A} = \pb{H, A}$. The equations of motion are easily integrated, we find
 \bea 
 {x}_\mu(\tau) &= X_\mu-{N} P_\mu \tau \label{xdot},
 \qquad 
 \chi_\mu(\tau) &= \lambda\left(A_\mu\cos\left(\frac{Ms}{\hbar} \tau\right) + B_\mu \sin\left(\frac{Ms}{\hbar} \tau\right)\right),
 \eea
 where $X_\mu,P_\mu,A_\mu$ and $B_\mu$ are constant vector solutions of 
 $P^2=-m^2$, $ A^2=B^2=1$ and $A\cdot P = B\cdot P=0$.
The momenta are simply given by
 \bea
 p_\mu = -\frac{\dot{x}_\mu}{N} = P_\mu ,\qquad 
 \pi_\mu = -\frac{\epsilon^2 \dot{\chi}_\mu}{M} .
 \eea
 This motion is expected, the coordinate $x_\mu$ evolves like a free particle while the dual coordinate $\chi_\mu$ undergoes oscillatory motion of frequency $M s/\hbar$ in the plane orthogonal to $P_\mu$. Furthermore, the motion is such that both orbital and spin angular momentum are constants of motion, specifically: $L_{\mu\nu} = (X\wedge P)_{\mu\nu}$ and $S_{\mu\nu} = {\hbar s}(A\wedge B)_{\mu\nu}$. 

\subsection*{Massive Second Order Formalism}
Further insights into the nature of DPS becomes apparent when we consider the second order formalism which is obtained from equation \eqref{dps-action} by integrating out the momenta and Lagrange multipliers. Only the main results will be presented here, for a more detailed analysis see Appendix \ref{app:second}. We begin by computing the equations of motion for the momenta and dual momenta which can be solved for $p_\mu$ and $\pi_\mu$ and then substituted back into the action, we find
\bea\label{action-2-main}
S&=& \int d\tau \left[\frac{\rho }{(N\tilde{N}-N_3^2)}- \frac{\tilde{M}}2({\chi^2} -\lambda^2)-\frac{N}2 m^2  + \frac{\tilde{N}}{2} \epsilon^2 s^2 \right],
\eea
where $\rho$ is given by
\bea
\rho&:=& \frac12 \left[ \tilde{N}(\dot{x}- N_4 \chi)^2 + N(\dot{\chi} -N_2 \chi)^2 -2 N_3(\dot{\chi} -N_2 \chi)\cdot(\dot{x}- N_4 \chi)
\right],
\eea
and we have introduce
\be
\tilde{N} = \frac{(M-N_1)}{\epsilon^2},\qquad 
\tilde{M} = \frac{s^2(M+N_1)}{\lambda^2}.
\ee
We can now solve for $N_2$ and $N_4$ which amounts to making the replacements
\begin{gather}
\dot{x}_\mu - N_4\chi_\mu \longrightarrow D_t  x_\mu:= \dot{x}_\mu - \frac{(\dot{ x}\cdot{\chi})}{\chi^2} \chi_\mu,\\
\dot{\chi}_\mu - N_2\chi_\mu \longrightarrow D_t \chi_\mu:= \dot{\chi}_\mu - \frac{(\dot{\chi}\cdot \chi)}{\chi^2} \chi_\mu,
\end{gather}
where $D_t$ is the time derivative projected orthogonal to $\chi$. It remains to integrate out the Lagrange multipliers $N$, $\tilde{N}$ and $N_3$; after some  algebra we obtain the following form for the action
\be\label{action-3-main}
{
S= \int \rd \tau \left[ \alpha\sqrt{ 
\epsilon^2 s^2 (D_t \chi)^2-m^2(D_t  x)^2  -
2  {s}\epsilon m \beta\left|(D_t  x)\wedge (D_t  \chi) \right|   } - \frac{\tilde{M}}2({\chi^2} -\lambda^2)
\right],}
\ee
where $\abs{(D_t x) \wedge (D_t \chi)} = \sqrt{(D_t x \cdot D_t\chi)^2 - (D_t x)^2(D_t \chi)^2}$ is a coupling between the particle motion and the spin motion, and $\alpha, \beta = \pm 1$ are signs used to define the square roots.
Observe that we can not integrate out the final Lagrange multiplier since the variation of $S$ with respect to $\tilde{M}$ is just the constraint $\chi^2 = \lam^2$. It can be checked that the momenta $p_x = \partial S/ \partial \dot{x}$ and $\pi_\chi = \partial S/ \partial \dot{\chi}$ satisfy the constraints
\begin{align}
  p_x^2 = -m^2, \quad   \pi_\chi^2 = \eps^2 s^2 ,\quad   \pi_\chi \cdot   \chi = 0, \quad   p_x \cdot   \pi_\chi = 0, \quad   p_x\cdot   \chi = 0.
\end{align}
Moreover when evaluated on--shell the action simplifies drastically and becomes
\begin{align}\label{action-simple-main}
S = \a \int d\tau \left\vert{m\abs{\dot{x}} - \beta \eps s \abs{\dot{\chi}}}\right\vert,
\end{align}
where we have defined $|\dot{x}| =\sqrt{-\dot{x}^2}$ and $\abs{\dot{\chi}} = \sqrt{\dot{\chi}^2}$. As expected, if $s = 0$ equation \eqref{action-simple-main} reduces to the action of a relativistic scalar particle. On the other hand, when $s \neq 0$ we can view the quantity appearing under the integral as the effective velocity of the particle. The effect of the the spin velocity $\dot{\chi}$ is seen to either decrease (for $\beta = +$) or increase (for $\beta = -$) this effective velocity relative to the scalar case.  Furthermore, given a trajectory $(x(t), \chi(t))$ we define the \textit{proper time} $\tau$ and the \textit{proper angle} $\theta$ as 
\begin{align}
\tau(t) := \int_0^t \abs{\dot{x}} dt', \qquad \theta (t) := \frac{1}{\lam}\int_0^t \abs{\dot{\chi}} d t'.
\end{align}
which can then be used to parametrize the motion
\begin{align}
x_\mu(t) = x_\mu - \frac{p_\mu}{m}\tau(t), \qquad \chi_\mu(t) = \chi_\mu \cos \theta(t) + \frac{\lam \pi_\mu}{\eps s}\sin \theta(t).
\end{align}

\subsection*{Massless Spinning Particle}
As mentioned earlier, a massless particle has four first class constraints, with $\Phi_3$ and $\Phi_4$ appearing in addition to $\Phi_{\rm{s}}$ and $\Phi_{\rm{m}}$, and so the relevant Hamiltonian is given by
\be
H = \frac{N}{2}p^2  + \frac{M}2 \left(\frac{\pi^2}{\epsilon^2} + \frac{s^2\chi^2}{\lambda^2} - 2s\right) + \frac{N_3}{\epsilon}({p\cdot\pi}) + \frac{s N_4}{\lambda}({p\cdot\chi}) .
\ee
Again the equations of motion are easily integrated, we find
\begin{align}
\chi^\mu (\tau) &= \lam\left(A^\mu  \cos\left(\frac{Ms}{\hbar}\tau\right) + B^\mu \sin\left(\frac{Ms}{\hbar}\tau\right) - \frac{N_4}{Ms}P^\mu \right),\\
x^\mu (\tau) &= X^\mu  + \tau\left(\frac{N_3^2+N_4^2}{M} - N\right)P_\mu + \frac{\eps }{M}\left(N_3\chi^\mu (t) + \frac{N_4\hbar}{Ms}\dot{\chi}^\mu (t)\right)\Bigg\vert_{t = 0}^{t = \tau},
\end{align}
where $X_\mu,P_\mu,A_\mu$ and $B_\mu$ are constant vector solutions of 
 $P^2=0$, $ A^2=B^2=1$ and $A\cdot P = B\cdot P=0$. The momenta are given by,
\begin{align}
p_\mu = P_\mu, \qquad \pi_\mu(\tau) = -\frac{\eps }{M}\left(\eps\dot{\chi}^\mu (\tau) + N_4 P_\mu\right).
\end{align}
Apart from a constant offset proportional to $P_\mu$ the evolution of $\pi_\mu$ and $\chi_\mu$ is identical to the massive particle. This is not the case for $x_\mu$ where, in addition to the expected linear evolution along $P_\mu$, there is oscillatory motion in the hyperplane orthogonal to $P_\mu$ of frequency $Ms/\hbar$ and amplitude $\hbar\sqrt{N_3^2 + N_4^2}/M$. This latter quantity, we note, is pure gauge, being a function of only the Lagrange multipliers $N_3$, $N_4$ and $M$.

\section{Coupling to Electromagnetism}
At this point DPS describes the free propagation of a relativistic spinning particle. Although our goal is to consider interactions between such particles it is important to show that DPS can be consistently coupled to electromagnetism. A coupling prescription is said to be consistent if it leaves the constraint structure invariant, lest the introduction of a background field fundamentally alter the system dynamics. \\
\indent At leading order we have the minimal coupling prescription
\begin{align}\label{minimal}
p_\mu \to P_\mu = p_\mu + eA_\mu(x),
\end{align}
which modifies the Poisson bracket of $P_\mu$ with itself $\pb{P_\mu, P_\nu} = -eF_{\mu\nu}$. Note that the pure spin constraints $\Phi_s$, $\Phi_1$ and $\Phi_2$ are unaffected by this adjustment. We can also include a higher order term via the spin orbit coupling $F_{\mu\nu}S^{\mu\nu}$ by making the replacement
\begin{align*}
\Phi_{\rm{m}}=\frac12(P^2+ m^2)  \to \Phi_{{\rm{m}},g} = \Phi_{\rm{m}} + \frac{eg}{4}F_{\mu\nu}{S}^{\mu\nu} ,
\end{align*}
where $g$ is the gyromagnetic ratio and $S_{\mu\nu}= (\chi\wedge \pi)_{\mu\nu}$ the spin bivector. These modifications alter the algebra of constraints which now reads
\bea
\{\Phi_{3}, \Phi_4 \} = -\left(P^2 -\frac{e}{2} F^{\mu\nu}{S_{\mu\nu}}\right)
=\widetilde{m}^2-2\Phi_{{\rm{m}},g},\\
\{\Phi_{{\rm{m}},g}, \Phi_3 \} = e (\pi_\mu  K^\mu )  ,\qquad
\{\Phi_{{\rm{m}},g}, \Phi_4 \} = e (\chi_\mu  K^\mu ).
\eea
where we have introduced an ``electromagnetic mass'' $\widetilde{m}$ and an  ``acceleration'' vector $K_	\mu$
\be
\widetilde{m}^2:= m^2+  \frac{e(g+1)}{2} F^{\mu\nu}{S_{\mu\nu}},\qquad 
K^\mu  := F^{\mu\nu} P_\nu - \frac{g}2 \left(  F^{\mu\nu} P_\nu -\frac{1}2 \partial ^\mu  F^{\nu\rho} S_{\nu\rho}\right). 
\ee
This vector enters the commutator
\be
\{ \Phi_{{\rm{m}},g}, P_\mu\}= e\left( K_\mu + \frac{ g}{2}  F_{\mu\nu} P^\nu  \right).
\ee

One can now check that, for a massive particle, this prescription does not change the number of degrees of freedom. The theory still possesses two first class and four second class constraints. In particular, $\Phi_{\rm{s}}$ remains first class since the spin sector is unmodified, while the other first class constraint is given by
\be
\Phi_{EM}:= \widetilde{m}^2 \Phi_{{\rm{m}},g} -e(\chi_\mu K^\mu ) \Phi_3 + e(\pi_\mu K^\mu ) \Phi_4.
\ee
The remaining four constraints will be second class and so the total Hamiltonian is given by
\be
H:= N \Phi_{EM} + M \Phi_{\rm s},
\ee
and it is straightforward to show that $H$ preserves all constraints. In standard phase space the resulting equations of motion are given by
\bea
\dot{x}^\mu &=& -N\left[\widetilde{m}^2 P^\mu 
+ e (SK)^\mu \right],\\
\dot{P}_\mu&=& Ne\left[ \widetilde{m}^2 \left( K_\mu + \frac{ g}{2}  (FP)_\mu \right)
+  e(FSK)_\mu\right].
 \eea
 where we have denoted $ (SK)^\nu = S^{\nu\rho}K_\rho$, $(FSK)_\mu = F_{\mu\nu}S^{\nu\rho} K_\rho$, etc... The equations of motion in dual phase space lead to\footnote{ They are explicitly given by
 \be
\dot{\chi}_\mu = -\frac{M}{\epsilon^2} \pi_\mu + eN 
\left(P_\mu K_\nu + \frac{g\widetilde{m}^2}2 F_{\mu\nu} \right)\chi^\nu ,\quad
\dot{\pi}_\mu = \frac{s^2M}{\lam^2} \chi_\mu  + eN\left(P_\mu K_\nu + \frac{g\widetilde{m}^2}2 F_{\mu\nu} \right)\pi^\nu .
 \ee}
\bea\label{frenkel-nyborg}
\dot{S}_{\mu\nu}
= Ne \left[
  P_\mu(SK)_\nu 
 +
\frac{g\widetilde{m}^2}2  (FS)_{\mu\nu} -(a\leftrightarrow b)
\right].
\eea
In the limit of weak ($\widetilde{m}^2 \approx m^2$) and constant electromagnetic field, equation \eqref{frenkel-nyborg} reduces to the Frenkel--Nyborg equation, \cite{frenkel_1926, nyborg_1961A}.\\
\indent For a massless particle, we can see that it is impossible to introduce an electromagnetic field while keeping $\Phi_3$ and $\Phi_4$ first class since their commutator involves the vector $K_\mu$. This means that  the minimal coupling prescription for a massless particle is inconsistent, it would change the number of degrees of freedom. This is hardly a surprise since it is well known that one cannot give  a photon  or a graviton an electromagnetic charge. 
 
\section{Interaction Vertex for Classical Spinning Particle}\label{sec:vertex}
We now come to the central result of this paper -- the interaction vertex for a relativistic spinning particle. In general, interactions between classical point particles are governed by a system of ten equations, conservation of linear momentum (four) and conservation of total angular momentum (six). The latter is represented in the DPS model by $J = x \wedge p + \chi \wedge \pi$ and is a constant of motion. For simplicity we restrict our attention to a trivalent vertex with one incoming and two outgoing particles, see Figure~\ref{fig:vertex}. The particles have phase space coordinates $(x_i,p_i),(\chi_i, \pi_i)$, $i = 1,2,3$ and so the conservation equations are given explicitly by 
\begin{gather}
p_1 = p_2 + p_3,\label{linear-mom}\\
(x_1\wedge p_1 + \chi_1 \wedge \pi_1) = (x_2\wedge p_2 + \chi_2\wedge \pi_2) + (x_3 \wedge p_3 + \chi_3 \wedge \pi_3).\label{total-mom}
\end{gather}
The coordinate $x_i$ denotes the spacetime location assigned to the interaction by particle $i$ and since one assumes that interactions are local in spacetime we should have that $x_1 = x_2 = x_3 = x$. Conservation of orbital angular momentum now follows immediately from locality and equation \eqref{linear-mom}; to be explicit
\begin{align}
x_1\wedge p_1 - x_2\wedge p_2 - x_3 \wedge p_3 = x\wedge (p_1 - p_2 - p_3) = 0. 
\end{align}
Thus, the system of equations we need to solve reduces to 
\begin{gather}
x_1=x_2=x_3=x ,\qquad p_1 = p_2 + p_3, \label{conserveP}\\
\chi_1 \wedge \pi_1  =  \chi_2 \wedge \pi_2 +  \chi_3 \wedge \pi_3 .\label{conserveS}
\end{gather}
Equation \eqref{conserveP} is standard, expressing the locality of interactions, which as mentioned in the introduction, goes hand in hand with conservation of the linear momentum. The second equation, which expresses conservation of spin angular momentum, requires some additional work to be properly interpreted. 
  
\subsection{Dual Locality}
We propose that conservation of spin angular momentum, equation \eqref{conserveS}, can be understood as an expression of the ``dual locality'' of the interaction vertex, i.e. interactions are ``local'' in dual phase space. Specifically, we assume that there exists a four--vector $\chi_\mu$ such that $\chi^2 = \lam^2$ and
\begin{align}\label{duallocal}
\chi_1 = \chi_2 = \chi_3 = \chi,
\end{align}
see Figure \ref{fig:vertex}. It follows from \eqref{conserveS} and \eqref{duallocal} that $\pi_1= \pi_2 + \pi_3 + \alpha \chi$ for some constant $\alpha$; contracting both sides with $\chi$ we get $\alpha \lambda^2 = \chi\cdot(\pi_1-\pi_2-\pi_3)$, the constraints $\chi_i \cdot \pi_i=\chi\cdot \pi_i = 0$ then imply $\chi$ is orthogonal to $\pi_i$ and so $\alpha = 0$. Thus, dual locality plus conservation of spin angular momentum intimates the conservation of dual momentum 
\begin{align}\label{consPI}
\pi_1 = \pi_2 + \pi_3.
\end{align}
This, we note, is an exact analogue of the results in standard phase space, further emphasising the duality of the dual phase space formulation.
\begin{figure}[t]
\begin{tikzpicture}
\node[fill=black, circle, scale = 0.3pt] (point) at (0,0) {};
\node[fill = none,below = 2cm of point, text = black!70!green] (income)  {$\scriptstyle p_1$};
\node[fill = none, above left = 2cm of point, text = red] (out1) {$\scriptstyle p_2$};;
\node[above right = 2cm of point, text = blue] (out2) {$\scriptstyle p_3$};;
\draw[->, thick, black!70!green] (income) to (point);
\draw [->, thick, red] (point) to (out1);
\draw[->, thick, blue] (point) to (out2);
\node[label = {[xshift =0.5cm, yshift = 1cm]$\scriptstyle \color{black!70!green} \chi_1,\pi_1$}, text = blue] at (income) {};
\node[label = {[xshift = 0.5cm, yshift = -1.6cm]$\scriptstyle \color{red} \chi_2,\pi_2$}, text = blue] at (out1) {};
\node[label = {[xshift = -0.5cm, yshift = -1.6cm]$\scriptstyle \color{blue} \chi_3,\pi_3$}, text = blue] at (out2) {};

\node[fill=black, circle, scale = 0.3pt] (pointDL) at (8,0) {};
\node[fill = none,below = 2cm of pointDL, text = black!70!green] (incomeDL)  {$\scriptstyle p_1$};
\node[fill = none, above left = 2cm of pointDL, text = red] (out1DL) {$\scriptstyle p_2$};;
\node[above right = 2cm of pointDL, text = blue] (out2DL) {$\scriptstyle p_3$};;
\draw[->, thick, black!70!green] (incomeDL) to (pointDL);
\draw [->, thick, red] (pointDL) to (out1DL);
\draw[->, thick, blue] (pointDL) to (out2DL);
\node[above = 0.1cm of pointDL]{$\scriptstyle \chi$};
\node[label = {[xshift = -0.3cm, yshift = 1cm]$\scriptstyle \color{black!70!green} \pi_1$}, text = blue] at (incomeDL) {};
\node[label = {[xshift = 0.6cm, yshift = -1.4cm]$\scriptstyle \color{red} \pi_2$}, text = blue] at (out1DL) {};
\node[label = {[xshift = -0.6cm, yshift = -1.4cm]$\scriptstyle \color{blue} \pi_3$}, text = blue] at (out2DL) {};

\draw [->, thick, black] (2.5,0) to (5.5,0);
\node[label = {[xshift = 0cm, yshift = 0.1cm]\small Dual Locality}, text = blue] at (4,0) {};
\end{tikzpicture}
\caption{Three particle interaction in DPS, with and without the assumption of dual locality}
\label{fig:vertex}
\end{figure}
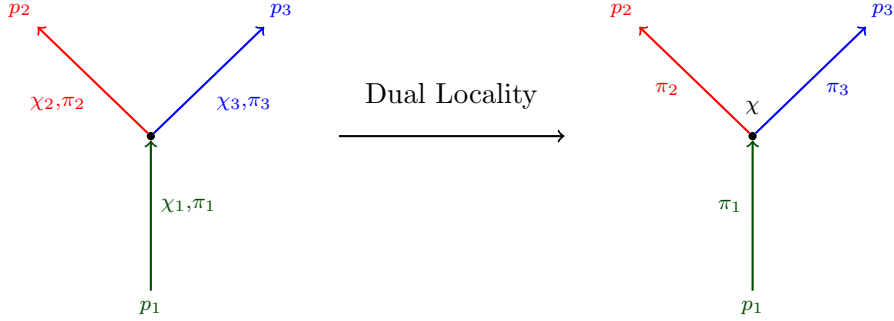\\
\indent To show that dual locality is a viable ansatz we must demonstrate that it is consistent with the constraints \eqref{DPconstraints}, which need to be satisfied for each particle and are enumerated below
\begin{enumerate}[i)]
\begin{minipage}[c]{0.25\textwidth}
\item $p_1\cdot \chi = 0$
\item $p_2\cdot \chi = 0$
\item $\chi^2 = \lam^2$
\end{minipage}
\begin{minipage}[c]{0.25\textwidth}
\item $p_1\cdot \pi_1 = 0$
\item $p_2 \cdot \pi_2 = 0$
\item $p_3\cdot \pi_3 = 0$
\end{minipage}
\begin{minipage}[c]{0.25\textwidth}
\item $\chi\cdot \pi_2 = 0$
\item $\chi \cdot \pi_3 = 0$
\item $\pi_1^2 = s_1^2$
\end{minipage}
\begin{minipage}[c]{0.25\textwidth}
\item $\pi_2^2 = s_2^2$
\item $\pi_3^2 = s_3^2$
\item $\pi_1 = \pi_2 + \pi_3$
\end{minipage}
\end{enumerate}
Notice that we have included conservation of dual momentum in this list, equation xii), since it will be convenient to have all restrictions on dual phase space variables collected in one spot. To proceed we use that conservation of momenta, equation \eqref{conserveP}, implies that $\lbrace p_1,p_2,p_3\rbrace$ span a two--plane, denoted $\mathsf{p}$. Introduce $\lbrace {e}_0, {e}_1\rbrace$ as an orthonormal basis for $\mathsf{p}$, where it is assumed that ${e}_0$ is time--like, we can then extend this to an orthonormal basis for $\R^4$ by including two additional vectors $\lbrace {e}_2,{e}_3\rbrace$. It will also be convenient to define a hodge dual in $\mathsf{p}$, 
denoted  
\begin{align}
(\tilde{ q})_\mu  := \eps_{\mu\nu\rho\sig}{e}_2^\nu{e}_3^\rho q^\sig,
\end{align}
for $q \in \mathsf{p}$.

\indent We now systematically solve the constraints beginning with i)-iii) which are easily seen to have solution
\begin{align}
\chi =\lambda( \cos\phi {e}_2 + \sin\phi {e}_3),
\end{align}
for some arbitrary angle $\phi$. Equations iv) -- vi) imply that the dual momenta $\pi_i$ lies in the hyper--plane orthogonal to $p_i$, hence we can expand $\pi_i$ as
\begin{align}
\pi_i =\alpha_i \tilde{p}_i + A_i {e}_2 + B_i {e}_3. \label{pi_i}
\end{align}
The hodge dual of \eqref{conserveP} implies $\tilde{p}_1 = \tilde{p}_2 + \tilde{p}_3$ and so, projecting xii) into the plane $\mathsf{p}$ and using \eqref{pi_i} gives
\begin{align}
(\alpha_1 - \alpha_2)\tilde{p}_2 + (\alpha_1 - \alpha_3)\tilde{p}_3 = 0.
\end{align}
Thus, if $p_2$ and $p_3$ are linearly independent we get $\alpha_1 = \alpha_2 = \a_3 = \a$. On the other hand, projecting xii) orthogonal to $\mathsf{p}$ and using \eqref{pi_i} again, requires $A_1 = A_2 + A_3$ and $B_1 = B_2 + B_3$.  Constraints vii) and viii) are then easily solved by putting $A_2 = -\beta \sin\phi$, $B_2 = \beta \cos\phi$ and $A_3 = -\gamma \sin\phi$, $B_3 = \gamma \cos\phi$ respectively. In summary we have
\begin{align}
\pi_1 &=\alpha\tilde{p}_1 + (\beta + \gamma)\chi^\perp\label{pi1},\\
\pi_2 &= \alpha\tilde{p}_2 + \beta \chi^\perp\label{pi2},\\
\pi_3 &=\alpha \tilde{p}_3 + \gamma \chi^\perp\label{pi3},
\end{align}
where $\chi^\perp = -\sin\phi {e}_2 + \cos\phi {e}_3$ is orthogonal to $\chi$. It remains to consider ix)--xi) which are seen to give
\begin{align}
m_1^2\a^2 + (\beta + \gamma)^2&= s_1^2,\label{bandc}\\
m_2^2\a^2 + \b^2 &= s_2^2, \label{b}\\
m_3^2\a^2 + \gamma^2 &= s_3^2.\label{c}
\end{align}
Before showing that the above equations possess a consistent solution we need to recall some restrictions on the mass and spin of the constituent particles, namely
\begin{gather}
m_2+ m_3\leq m_1,\label{mass}\\
\abs{s_2-s_3} \leq s_1 \leq s_2 + s_3.\label{spin}
\end{gather}
The first inequality is well known, and easily derived from momentum conservation \eqref{conserveP}. Equation \eqref{spin} on the other hand is a quantum mechanical result derived by considering the eigenvalues of the total angular momentum operator in a composite system. Here we will show that it follows from the assumption of dual locality. We begin by squaring equation \eqref{consPI} to obtain 
\begin{align*}
s_1^2 = s_2^2 + s_3^2 + 2\pi_2\cdot \pi_3.
\end{align*}
As $\pi_i$ is spacelike we can apply the Cauchy--Schwartz inequality with impunity
\begin{align*}
2|\pi_2\cdot\pi_3| \leq 2 \abs{\pi_2}\abs{\pi_3} = 2s_2s_3. 
\end{align*}
Substituting this result into the previous equation gives $(s_2-s_3)^2 \leq s_1^2 \leq (s_2+s_3)^2$, and the desired result follows after taking square roots.\\
\indent With this in mind we return to equations \eqref{bandc}-\eqref{c}. The latter two can be used to solve for $\beta$ and $\gamma$ in--terms of $\a$ and the result substituted into \eqref{bandc}. After rearranging and taking the square we get a consistency condition for $\a$
\be\label{alpha-cond}
 (s_2^2- m_2^2 \alpha^2)(s_3^2- m_3^2 \alpha^2) = (S^2 -M^2\a^2)^2,
\ee
where $2M^2 := m_1^2-m_2^2 - m_3^2$ and $2S^2 := s_1^2 - s_2^2 - s_3^2$. It is not enough to simply solve this equation for $\a$ since it is immediately obvious from \eqref{bandc} - \eqref{c} that $\a^2 \leq r^2_i$ where $r_i = s_i/m_i$ for $m_i \neq 0$. As such, we introduce variables $\theta_2$ and $\theta_3$ which satisfy
\begin{align}\label{restrict}
\alpha= r_2\cos\theta_2 = r_3\cos\theta_3,
\end{align}
and without loss of generality suppose $r_3 \leq r_2$. Note that we can choose the sign of $\theta_2$ and $\theta_3$ so that $\beta = s_2 \sin \theta_2$ and $\gamma = s_3 \sin \theta_3$. The consistency equation on $\alpha$ now reads  $F(\theta_3)=0$ where
\be\label{lasteq}
F(\theta) := (S^2- M^2 r_3^2 \cos^2 \theta)^2  -  s_2^2 s_3^2 \sin^2\theta
\left(1-\tfrac{r_3^2}{r_2^2} + \left(\tfrac{r_3}{r_2} \sin \theta\right)^2\right).
\ee
It suffices, therefore, to show that $F(\theta)$ has a zero in the interval $[-\pi/2, \pi/2]$ and so we note that 
\begin{align*}
F(0) = (S^2- M^2 r_3^2 )^2\geq 0,\qquad F(\pm \pi/2) = -[ (s_2 + s_3)^2-s_1^2 ][s_1^2 - (s_2 - s_3)^2] \leq 0,
\end{align*}
where the second equality follows from \eqref{spin}. By the intermediate value theorem there exists $\bar{\theta} \in [0, \pi/2]$ such that $F(\pm \bar{\theta}) = 0$ and so $\a = r_3\cos\bar{\theta}$ satisfies \eqref{alpha-cond}. It follows, for massive particles, that there are two solutions to the dual locality equations for which $\alpha>0$. These two solutions are related by a change of orientation in the plane orthogonal to $\mathsf{p}$; if $(\alpha,\beta,\gamma)$ is a solution then
$(\alpha,-\beta,-\gamma)$ is also a solution. Note that by parity invariance 
$(-\alpha,-\beta,-\gamma)$ and $(-\alpha,\beta, \gamma)$ are also  solutions.\\
\indent The case where $m_2 = 0$ can be obtained from the above by  allowing $r_2 \to \infty$ in \eqref{lasteq}, and one can again obtain a solution for $\a$ by using the intermediate value theorem. In the remaining case\footnote{It is impossible to have three massless interacting particles.}  $m_2 = m_3 = 0$, equations \eqref{b} and \eqref{c} are solved immediately as $\beta = \eps_2 s_2$ and $\gamma = \eps_3s_3$ where $\eps_i = \pm 1$. We then obtain for $\alpha$
\begin{align*}
\alpha^2 = \frac{1}{ m_1^2}\left(s_1^2 - (\eps_2 s_2 + \eps_3s_3)^2\right),
\end{align*}
where equation \eqref{spin} implies that $\eps_2\eps_3 = -1$ and we again find four solutions belonging to two sectors related by parity. This completes our analysis of the three particle interaction, showing that dual locality ensures a consistent vertex for any viable combination of spinning particles.

\subsection{Universality of Dual Locality}
The previous section established dual locality as a sufficient condition to ensure a consistent three--point vertex, we now establish its necessity. 
The key point is that when the spin is non--zero, we have an additional gauge symmetry in the system which corresponds to a rotation in the $(\chi,\pi)$ plane, recall equation \eqref{gauge}:
\be
R_\theta(\chi_\mu,\pi_\mu) =( \cos \theta \chi_\mu + \tfrac{\lambda}{\epsilon s} \sin \theta \pi_\mu, \cos \theta \pi_\mu - \tfrac{\epsilon s}{\lambda} \sin \theta \chi_\mu).
\ee
Such a gauge transformation does not change the value of the spin bivector
$ R_\theta(\chi) \wedge R_\theta(\pi) = \chi \wedge \pi$. Therefore if 
$(\chi_i,\pi_i)_{i=1,2,3}$ is a solution of (\ref{conserveS}) then 
$ (R_{\theta_i}(\chi_i),R_{\theta_i}(\pi_i))_{i=1,2,3}$ is also a solution for arbitrary $\theta_i$.
This is simply an expression of the gauge symmetry of the theory.
The main claim we now want to prove is that {\it any} solution of the spin conservation equation (\ref{conserveS}) is gauge equivalent to a solution satisfying dual locality.
In other words if  $(\chi_i,\pi_i)_{i=1,2,3}$ is a solution of (\ref{conserveS}) then there exists  $(\chi',\pi_i')_{i=1,2,3}$ with $\pi_1' =\pi_2'+\pi_3'$, and $\theta_i$  such that 
\be
(\chi_i,\pi_i) = (R_{\theta_i}(\chi'),R_{\theta_i}(\pi_i')), \,\, \mathrm{for} \,\, {i=1,2,3}.
\ee
Note that in addition to the rotation \eqref{gauge} DPS is invariant under the global re--scaling $\lam \to \a \lam$ and $\eps \to \a^{-1} \lam$. Therefore, we can assume that all $\lam_i$ and $\eps_i$ have been re--scaled to some common values $\lam$ and $\eps$.\\
\indent Suppose that we have a solution to equations \eqref{conserveP} and \eqref{conserveS}, including all accompanying constraints. It is always possible to choose $\chi_i$ orthogonal to the plane $\mathsf{p}$, to see why consider $\chi_2$: By construction $\chi_2 \cdot  p_2 = 0$ and so we need only ensure that it is orthogonal to $p_3$ since then conservation of momentum guarantees that it will be orthogonal to $p_1$ as well. Hence, if $\chi_2 \cdot p_3 \neq 0$ a gauge rotation with $\cot\theta = \lambda \pi_2\cdot p_3/(s_2\chi_2\cdot p_3)$, will ensure that the new $\chi_2$ is orthogonal to $p_3$. A similar argument holds for the other $\chi_i$ and the claim is justified thereby allowing us to write $\chi_i = \lambda (\cos\phi_i {e}_2 + \sin\phi_i {e}_3)$, since $\chi_i^2 =\lambda^2$.  Now, contract \eqref{conserveS} with $(p_1,p_2,p_3)$ to obtain
\begin{align}\label{proof1}
 \chi_2(p_3\cdot \pi_2) + \chi_3(p_2\cdot \pi_3) &= 0,\\
 \chi_1(p_2\cdot \pi_1) - \chi_3(p_2\cdot \pi_3) &= 0,\label{proof2}\\
 \chi_1(p_3\cdot \pi_1) - \chi_2(p_3\cdot \pi_2) &= 0. \label{proof3}
\end{align}
There are two cases to consider.  Either  $(p_i\cdot \pi_j)_{i\neq j}$ are all vanishing or they are all  non--vanishing.
Indeed, if $p_3\cdot \pi_2 = 0$ the above equations imply that $p_2\cdot \pi_3=p_2\cdot \pi_1 = p_3 \cdot \pi_1 = 0$ which in turn, via momentum conservation, yield $p_1\cdot \pi_2=p_1\cdot\pi_3=0.$ \\
\indent Let us  first  assume  that $p_i\cdot \pi_j = 0$. As argued above, $\chi_i$ and $\pi_i$ are orthogonal to $\mathsf{p}$ and therefore can be expanded as
\begin{align*}
\chi_i = \lambda(\cos\phi_i {e}_2 + \sin\phi_i {e}_3),\qquad  \pi_i =s_i\eps( -\sin\phi_i  {e}_2 + \cos\phi_i {e}_3).
\end{align*}
A further gauge transformation with $\theta_i = -\phi_i$ can now be performed to give $\chi_i= \lambda e_2$, $\pi_i = s_i \eps e_3$, which proves the proposition.
 \\
\indent In the generic case we have $(p_i\cdot \pi_j)_{i\neq j} \neq 0$. Contract equation \eqref{proof1} with $\pi_3$ to obtain $\pi_3\cdot \chi_2 = 0$; repeating for the other $\pi_i$ we find that $(\chi_i)_{i=1,2,3}$ is orthogonal to $(\pi_j)_{i=1,2,3}$. 
With this established we can return  to equation \eqref{conserveS} contract with $\chi_1$ and then $\chi_2$ and combine the results to eliminate the terms proportional to $\pi_1$, we find
\begin{align*}
0 = \left[(\chi_1\cdot\chi_2)^2 -\lambda^4 \right]\pi_2 + \left[(\chi_1\cdot\chi_3)(\chi_1\cdot \chi_2) - \lambda^2(\chi_2\cdot\chi_3)\right]\pi_3.
\end{align*}
Note that $\pi_2$ and $\pi_3$ can not be parallel since then $\pi_2 \cdot p_3 \propto \pi_3\cdot p_3 = 0$ which is contrary to the original assumption $\pi_2\cdot p_3 \neq 0$. Hence, the previous equation implies that
\begin{align*}
|\chi_1\cdot \chi_2| = \lambda^2.
\end{align*}
As $\chi_i$ are space--like vectors which satisfy $\chi_i^2 = \lambda^2$, the Cauchy--Schwartz inequality implies that $\chi_1$ and $\chi_2$ are parallel, hence  $\chi_1 = \pm \chi_2$.  We can repeat the above procedure, contracting \eqref{conserveS} with $\chi_1$ and $\chi_3$, to obtain $\chi_1 = \pm \chi_3$, and so 
\begin{align*}
\epsilon_1\chi_1 = \epsilon_2\chi_2 = \epsilon_3 \chi_3 = \chi,
\end{align*}
where $\eps_i = \pm$. This is not exactly what we want. All we have to do is perform another set of gauge transformations by the angle $(1-\eps_i)\pi/2$ to  transform $ (\chi_i,\pi_i)\to (\epsilon_i\chi_i,\epsilon_i \pi_i)$. Note that these gauge transformations do not affect any of the orthogonality properties established before and so we obtain the dual locality property
\be
\chi_1 = \chi_2 =  \chi_3 = \chi,\qquad \pi_1=\pi_2+\pi_3.
\ee
This completes the proof, showing that a solution to \eqref{conserveP} and \eqref{conserveS} implies that dual locality holds, up to a gauge re--labelling.

\subsection{An Alternative View of Dual Locality}
The universality of dual locality is an important result further emphasising the symmetry between standard and dual phase space. As such, it will be beneficial to see how dual locality arises from one of the alternative models presented earlier in this paper. In particular, select the parametrization of Section \ref{sec:vectsphere}, where spin is represented by a single vector $n_\mu$. Recall that $n_\mu$ has the interpretation of an $S^2$ vector boosted in the direction of the particles momenta and the spinning part of angular momentum is given by $S_{\mu\nu} = s*(n \wedge u)_{\mu\nu}$. Consider again a three particle interaction with one particle incoming and the other two out going. In what follows we will assume $m \neq 0$.
\begin{figure}[h]\label{fig:sphere}
\begin{tikzpicture}
	\shade[ball color=blue!80!green,opacity=0.50] (0,0) circle (5cm);
	\draw[thin, draw opacity = 0.7] (0,0) circle (5cm);
    
    \draw[thin, fill = orange,opacity = 0.3, draw opacity = 0.7]  (-5,0) arc (180:360:5cm and 0.5cm);
    \draw[thin,fill = orange ,opacity = 0.3, draw opacity = 0.7,dashed] (-5,0) arc (180:0:5cm and 0.5cm);
%
    \draw[thin,fill = green,opacity = 0.3, draw opacity = 0.7]  (0,5) arc (90:-90:2.5cm and 5cm);       
    \draw[thin,fill = green,opacity = 0.3, draw opacity = 0.7, dashed] (0,-5) arc (-90:-270:2.5cm and 5cm);

    \draw[gray,draw opacity = 0.7,->](0,0) -- (0,7);
	\draw [gray,draw opacity = 0.7,->](0,0)-- (7,0);
	\draw[gray,draw opacity = 0.7,->] (0,0) -- (-4.95,-4.95);
	
   \draw[->, yellow] (0,0)-- node[ label={[label distance = 0.1cm]above: $n_1$}]{}(2.45,1);
    \draw[densely dashed, yellow, draw opacity = 0.7] (2.45,1) -- (2.45,-0.45);
    \draw[densely dashed,yellow,draw opacity = 0.7] (0,0) -- (2.45,-0.45);

	\draw[->, red] (0,0)-- (-1.6,3.85)node[font = \tiny,label={[label distance = 0.2cm]south east: $\hat{\pi}_1$}]{};
  	\draw[densely dashed, red,draw opacity = 0.7 ] (-1.6,3.85) -- (-1.6,0.29);
    \draw[densely dashed, red,draw opacity = 0.7] (0,0) -- (-1.6,0.29);

	\draw [->, blue] (0,0) -- (-1.05,-0.50)node[font = \tiny, label={[label distance = -0.3cm]195: $\hat{\chi}$}]{};
    
\end{tikzpicture}
\caption{Relationship between $\chi$ and $\pi$ and $n$ for particle one, plotted in the hyperplane orthogonal to $p_1$. To ensure that all vectors sit on the same sphere we have plotted the unit vectors $\hat{\chi}$ and $\hat{\pi}_1$.}
\label{fig:rotsphere}
\end{figure}
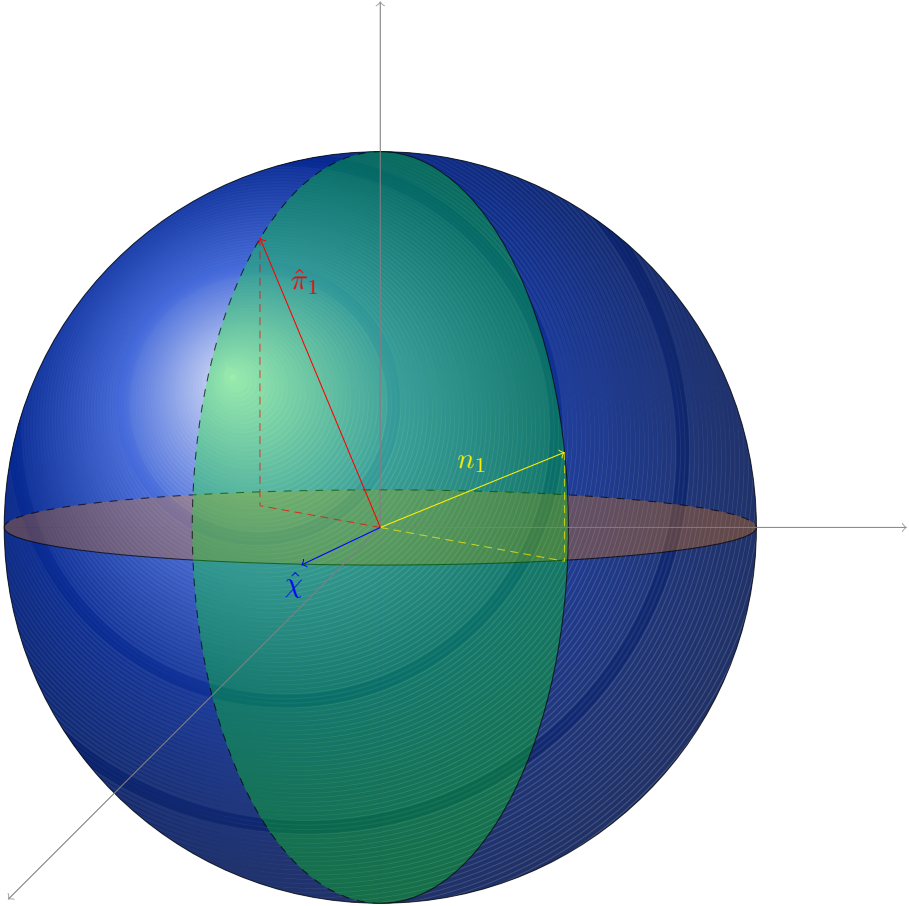
\\\indent Interactions, as previously discussed, are governed by conservation of linear momentum, equation \eqref{conserveP}, and conservation of spin angular momentum. The latter, after taking the hodge dual and making use of \eqref{conserveP}, can be written as  
\begin{align}\label{zeta}
[(r_1n_1 - r_2n_2) \wedge p_2] + [(r_1n_1 - r_3n_2) \wedge p_3] = 0,
\end{align}
where $ r_i =s_i/m_i$.
Let $A^{\perp_\mathsf{p}}$ denote the projection of a vector $A$ onto the plane orthogonal to $\mathsf{p}$, applying this projection to \eqref{zeta} yields\footnote{Assuming $p_2$ and $p_3$ are linearly independent.}
\begin{align}\label{locality}
r_1n_1^{\perp_\mathsf{p}} = r_2n_2^{\perp_\mathsf{p}} = r_3n_3^{\perp_\mathsf{p}}.
\end{align}
A further condition on the $n_i$ is obtained by contracting \eqref{zeta} with $\tilde{p}_2\wedge \tilde{p}_3$, viz
\begin{align}\label{zetap}
r_1n_1\cdot \tilde{p}_1 = r_2n_2\cdot \tilde{p}_2  + r_3n_3\cdot \tilde{p}_3.
\end{align}
The previous two equations provide a natural method for defining variables $\chi$ and $\pi_i$ which satisfy dual locality, in particular
\begin{align*}
\chi = \frac{\lam}{\abs{n_i^{\perp_\mathsf{p}}}} *(e_0 \wedge e_1 \wedge n_i) \qquad \mr{and} \qquad \pi_i = \frac{\eps r_i}{\lam} *(\chi \wedge p_i \wedge n_i).
\end{align*}
It follows from \eqref{locality} that $\chi$ is independent of $i$ while \eqref{zetap} can be used to show $\pi_1 = \pi_2 + \pi_3$. The necessary constraints, see i) -- xi), are also satisfied as one can easily check. Note that the above definitions are ambiguous up to a sign, although the same sign must be chosen for all $\pi_i$, and so we see again that there are four possible solutions belonging to two parity related sectors. In summary, conservation of angular momentum requires that the vectors $r_in_i$ are equal when projected into the plane $\mathsf{p}^\perp$. The dual position $\chi$ is then the unique (up to a sign) vector of length $\lam$, lying in the plane $\mathsf{p}^\perp$ which is orthogonal to $r_i n_i^{\perp_\mathsf{p}}$. In turn, the dual momenta $\pi_i$ is the unique (up to a sign) vector of length $\eps s_i$ orthogonal to $p_i$, $n_i$ and $\chi$. We have pictured the relationship between $\chi$, $\pi_1$ and $n_1$ in Figure \ref{fig:rotsphere}, with the sphere sitting in the hyper--plane orthogonal to $p_1$. The projections of $p_2$ and $p_3$ into this hyper--plane are parallel (or anti--parallel) and define a normal vector for the equatorial plane, which is identified with $\mathsf{p}^\perp$.

\section{Conclusion}
The dual phase space model presented herein provides a unique perspective on the relativistic spinning particle. It imagines the variables which parametrize the spinning degrees of freedom as living in their own phase space, independent from the standard phase space of $x_\mu$ and $p_\mu$. The structure of these spaces and their relationship with one another are governed by constraints which yield linear propagation of the coordinate variable as well as oscillatory motion of the dual variables. Furthermore, the on--shell action for DPS is seen to be the sum of the usual proper time which encodes motion in spacetime, and a ``proper angle'' determined by the spin motion.\\
\indent The three point interaction vertex is further testament to the utility of DPS. Given that interactions are governed by conservation of linear and angular momentum we showed that consistency is possible if and only if dual locality is satisfied, in which case conservation of angular momentum becomes conservation of dual momenta. This further emphasises the symmetry between the two phase spaces and provides a prescription for implementing interactions in the worldline formalism. \\
\indent If, for a moment, one ignores the constraints then DPS implies that the phase space of a spinning particle is \textit{twice} as large as that of a scalar particle. If one takes the duality of DPS seriously, then this doubling suggests that one might be able to realize the spinning particle as a composite of two scalar particles. The behaviour of the spinning particle itself could then be viewed as emerging from interactions between the constituent particles. This is a radical proposal which we intend to explore more fully in a subsequent paper.

\appendix \section{Second Order Formulation}\label{app:second}
In this appendix we present the second order formulation of the DPS action which is obtained from equation \eqref{dps-action} by integrating out the momenta and all Lagrange multipliers. We begin by re--writing the action as
\begin{align}\label{action-1}
S &= \int d\tau 
\left[
p_\mu(\dot{x}^\mu - N_4 \chi^\mu ) 
+ \pi_\mu(\dot{\chi}^\mu - N_2\chi^\mu ) 
 - \frac{N}{2}(p^2 +m^2) 
 - \frac{\tilde{N}}{2} ({\pi^2} -\epsilon^2s^2)
 -  N_3({p\cdot \pi})
\right.\\ 
& \qquad \quad\left. - \frac{\tilde{M}}{2}({\chi^2} -\lambda^2) \right].
\end{align}
where we have introduced 
\be
\tilde{N} = \frac{(M-N_1)}{\epsilon^2},\qquad 
\tilde{M} = \frac{s^2(M+N_1)}{\lambda^2}.
\ee
The equations of motion for the momenta read
\bea
 N {p_\mu} + N_3{\pi_\mu} &=& (\dot{x}_\mu- N_4 \chi_\mu), \\
 N_3 {p_\mu} + \tilde{N}  \pi_\mu &=&(\dot{\chi}_\mu -N_2 \chi_\mu),
\eea 
and upon inverting these we obtain
\bea \label{momenta}
 (N\tilde{N}-N_3^2)  {p_\mu}  &=& \tilde{N}(\dot{x}_\mu- N_4 \chi_\mu)-N_3(\dot{\chi}_\mu -N_2 \chi_\mu), \\
 (N\tilde{N}-N_3^2)  \pi_\mu &=&-{N}_3(\dot{x}_\mu- N_4 \chi_\mu)+ N(\dot{\chi}_\mu -N_2 \chi_\mu).
\eea
Substituting this result into \eqref{action-1} we find 
\bea\label{action-2}
S&=& \int d\tau \left[\frac{\rho }{(N\tilde{N}-N_3^2)}- \frac{\tilde{M}}2({\chi^2} -\lambda^2)-\frac{N}2 m^2  + \frac{\tilde{N}}{2} \epsilon^2 s^2 \right],
\eea
where $\rho$ is given by
\bea
\rho&:=& \frac12 \left[ \tilde{N}(\dot{x}- N_4 \chi)^2 + N(\dot{\chi} -N_2 \chi)^2 -2 N_3(\dot{\chi} -N_2 \chi)\cdot(\dot{x}- N_4 \chi)
\right].
\eea
We can now start integrating out the constraints, begin by varying \eqref{action-2} with respect to $N_2$ and $N_4$, then
\be
N_2 \chi^2 = \dot{\chi}\cdot \chi,\qquad 
N_4 \chi^2 = \dot{x}\cdot\chi.
\ee
This suggests that we introduce the notation
\be
D_t  x_\mu:= \dot{x}_\mu - \frac{(\dot{ x}\cdot{\chi})}{\chi^2} \chi_\mu,
\qquad 
D_t \chi_\mu:= \dot{\chi}_\mu - \frac{(\dot{\chi}\cdot \chi)}{\chi^2} \chi_\mu,
\ee
where $D_t$ is the time derivative projected orthogonal to $\chi$. We can now compute the variation with respect to the Lagrange multipliers $N,\tilde{N}$ and $N_3$; after some algebra we find
\bea
{(D_t \chi)^2 }  
&=& {\tilde{N}^2 \epsilon^2s^2 - N_3^2 m^2},\label{tN}\\
{(D_t  x)^2 }&=&
{{N}^2_3 \epsilon^2s^2 - N^2 m^2}, \label{N}\\
{(D_t \chi)\cdot(D_t x) } &=& {{N_3}\tilde{N} \epsilon^2s^2 - N_3N m^2 }. \label{N4}
\eea 
To solve for these equations it will be convenient to define
\bea
D:= (N\tilde{N}-N_3^2) {s} \epsilon m,\qquad T:= (\tilde{N} \epsilon^2s^2-N m^2),
\eea
which allow us to re--write \eqref{tN}-\eqref{N4} as
\bea\label{solved}
{(D_t \chi)^2 }  
= \tilde{N} T +  \tfrac{mD}{{s} \epsilon} ,\qquad
{(D_t \chi)\cdot(D_t  x) } = {{N_3} T  },\qquad
{(D_t x)^2 }&=& NT -  \tfrac{{s} \epsilon D}{m}. 
\eea 
These relations are straightforward to invert and we find
\bea
D &=& \beta \sqrt{
\left[(D_t \chi)\cdot(D_t  x)\right]^2-
(D_t  x)^2(D_t \chi)^2 }= \beta \left|(D_t  x)\wedge (D_t  \chi) \right|,\label{d-solved}\\
T&=& \alpha \sqrt{ 
\epsilon^2 s^2 (D_t \chi)^2-m^2(D_t  x)^2  -
2\beta {s} \epsilon m \left|(D_t  x)\wedge (D_t  \chi) \right|  },\label{t-solved}
\eea
where $\alpha=\pm1$ and $\beta=\pm1 $ are signs needed to define the square root. 
For definiteness we choose both signs to be positive from now on. Thus, after integration of $N_2,N_4$ and $N,\tilde{N}$ and $N_3$ the action becomes 
\be\label{action-3}
{
S= \int \rd \tau \left[ \alpha\sqrt{ 
\epsilon^2 s^2 (D_t \chi)^2-m^2(D_t  x)^2  -
2  {s}\epsilon m \beta\left|(D_t  x)\wedge (D_t  \chi) \right|   } - \frac{\tilde{M}}2({\chi^2} -\lambda^2)
\right].}
\ee
Observe that we can not integrate out the final Lagrange multiplier since the variation of $S$ with respect to $\tilde{M}$ is just the constraint $\chi^2 = \lam^2$. We can however, obtain expressions for some of the other Lagrange multipliers, viz
\begin{align}
N &= \frac{m({D}_t x)^2 + s\eps\beta \abs{({D}_t x) \wedge ({D}_t \chi)}}{mT},\\
\tilde{N} &= \frac{s\eps({D}_t \chi)^2  -  m \beta \abs{({D}_t x) \wedge ({D}_t \chi)}}{s\eps T},\\
N_3 &= \frac{ [({D}_t x)\cdot ({D}_t \chi)]}{T}.
\end{align}  
The conjugate momenta are now obtained via the standard prescription $p_x = \partial S/ \partial \dot{x}$ and $\pi_\chi = \partial S / \partial \dot{\chi}$, we find
\begin{alignat}{4}
{p}_{x,\mu} &=& -\frac{m}{T}\left(m {D}_t   x_\mu + \frac{\beta s\eps}{\abs{{D}_t   x \wedge {D}_t  \chi}} \left[({D}_t  x \cdot {D}_t  \chi) {D}_t  \chi_\mu - ({D}_t  \chi)^2 {D}_t  x_\mu \right]\right),\\
\pi_{\chi,\mu}   &=&  \frac{s\eps}{T}\left(s\eps {D}_t  \chi_\mu - \frac{m\beta }{\abs{{D}_t   x \wedge {D}_t  \chi}} \left[({D}_t  x \cdot{D}_t  \chi) {D}_t  x_\mu - ({D}_t  x)^2 {D}_t  \chi_\mu \right]\right).
\end{alignat}
It can be checked that these momenta satisfy the constraints
\begin{align}
  p_x^2 = -m^2, \quad   \pi_\chi^2 = \eps^2 s^2 ,\quad   \pi_\chi \cdot   \chi = 0, \quad   p_x \cdot   \pi_\chi = 0, \quad   p_x\cdot   \chi = 0.
\end{align}
\indent The variation of the action with respect to $x_\mu$ and $\chi_\mu$ determines the Lagrange equations of motion, in particular
\begin{align}
\dot{p}_{x,\mu} = 0, \qquad \dot{\pi}_{\chi,\mu} =  -(  \chi \cdot \dot{x})  p_{x,\mu} - (  \chi \cdot \dot{\chi})  \pi_{\chi,\mu} -\tilde{M}\chi_\mu.
\end{align}
Provided we implement $\chi^2 = \lam^2$ these equations preserve $p_x^2 = -m^2$ and  $\pi_\chi^2 = \eps^2 s^2$; demanding that $\pi_\chi \cdot \chi = 0$ is also preserved in time determines the Lagrange multiplier $\tilde{M}$:
\begin{align}
\tilde{M} = \frac{\eps^2s^2}{\lam^2}\tilde{N}.
\end{align} 
On the other hand, for the remaining two constraints we have
\begin{align}
\frac{d}{dt} (p_x\cdot \chi) = -\frac{m^2}{T}({D}_t x) \cdot ({D}_t \chi), \qquad \frac{d}{dt}(p_x \cdot \pi_\chi) = m^2(\chi\cdot \dot{x}).
\end{align}
Therefore, ensuring that these quantities are stationary in time requires that we impose constraints on the initial conditions, specifically $({D}_t  {\chi})\cdot({D}_t  {x}) =\dot{x}\cdot\chi = 0$. These are equivalent, when $\chi^2 = \lam^2$, to $\dot{  x}\cdot  \chi=\dot{  x}\cdot \dot{ \chi}=0$ which implies that the dual motion is always orthogonal to the particle velocity. Once these extra constraints are imposed, the action simplifies to the one quoted in the main text, see equation \eqref{action-simple-main}
\begin{align}\label{action-simple}
S = \a \int d\tau \left\vert{m\abs{\dot{x}} - \beta \eps s \abs{\dot{\chi}}}\right\vert,
\end{align}
where we have defined $|\dot{x}| =\sqrt{-\dot{x}^2}$ and $\abs{\dot{\chi}} = \sqrt{\dot{\chi}^2}$.

\section{Dirac Brackets}\label{app:dirac-brackets}
\noindent We include here an explicit formulation of the Dirac brackets for DPS. Assuming $m \neq 0$ a direct computation gives
\begin{equation}
\begin{aligned}
\DB{f,g} &= \pb{f,g} + \frac{1}{2s^2}\left(\pb{f,\Phi_1}\pb{\Phi_2,g} - \pb{f, \Phi_2}\pb{\Phi_1,g}\right)  \\
&\qquad + \frac{1}{m^2}\left(\pb{f,\Phi_3}\pb{\Phi_4,g} - \pb{f,\Phi_4}\pb{\Phi_3,g}\right).
\end{aligned}
\end{equation}
The commutation relations between the phase space variables are now given by
\begin{align}
\DB{x^\mu, p^\nu} &= \eta^{\mu\nu}, \quad & \DB{x^\mu, x^\nu} &= \frac{1}{m^2}(\chi \wedge \pi)^{\mu\nu}, \\
\DB{x^\mu, \chi^\nu} &= \frac{1}{m^2}\chi^\mu p^\nu, \quad & \DB{\chi^\mu, \chi^\nu} &= -\frac{1}{2\eps^2 s^2}\left(\chi\wedge \pi\right)^{\mu\nu},\\
\DB{x^\mu, \pi^\nu} &= \frac{1}{m^2}\pi^\mu p^\nu, \quad &  \DB{\pi^\mu, \pi^\nu}&= -\frac{s^2}{2\lam^2}(\chi \wedge \pi)^{\mu\nu},\\
&\;   \DB{\chi^\mu, \pi^\nu} = \eta^{\mu\nu} - \frac{s}{2\lam^2}\chi^\mu\chi^\nu - \frac{1}{2\eps^2 s^2}\pi^\mu\pi^\nu + \frac{1}{m^2}p^\mu p^\nu. \hspace{-3cm}
\end{align}      
To obtain the brackets for a massless particle let $m \to \infty$ in the above relations. 

\bibliography{spinref}{}

\begin{thebibliography}{64}%
\makeatletter
\providecommand \@ifxundefined [1]{%
 \@ifx{#1\undefined}
}%
\providecommand \@ifnum [1]{%
 \ifnum #1\expandafter \@firstoftwo
 \else \expandafter \@secondoftwo
 \fi
}%
\providecommand \@ifx [1]{%
 \ifx #1\expandafter \@firstoftwo
 \else \expandafter \@secondoftwo
 \fi
}%
\providecommand \natexlab [1]{#1}%
\providecommand \enquote  [1]{``#1''}%
\providecommand \bibnamefont  [1]{#1}%
\providecommand \bibfnamefont [1]{#1}%
\providecommand \citenamefont [1]{#1}%
\providecommand \href@noop [0]{\@secondoftwo}%
\providecommand \href [0]{\begingroup \@sanitize@url \@href}%
\providecommand \@href[1]{\@@startlink{#1}\@@href}%
\providecommand \@@href[1]{\endgroup#1\@@endlink}%
\providecommand \@sanitize@url [0]{\catcode `\\12\catcode `\$12\catcode
  `\&12\catcode `\#12\catcode `\^12\catcode `\_12\catcode `\%12\relax}%
\providecommand \@@startlink[1]{}%
\providecommand \@@endlink[0]{}%
\providecommand \url  [0]{\begingroup\@sanitize@url \@url }%
\providecommand \@url [1]{\endgroup\@href {#1}{\urlprefix }}%
\providecommand \urlprefix  [0]{URL }%
\providecommand \Eprint [0]{\href }%
\providecommand \doibase [0]{http://dx.doi.org/}%
\providecommand \selectlanguage [0]{\@gobble}%
\providecommand \bibinfo  [0]{\@secondoftwo}%
\providecommand \bibfield  [0]{\@secondoftwo}%
\providecommand \translation [1]{[#1]}%
\providecommand \BibitemOpen [0]{}%
\providecommand \bibitemStop [0]{}%
\providecommand \bibitemNoStop [0]{.\EOS\space}%
\providecommand \EOS [0]{\spacefactor3000\relax}%
\providecommand \BibitemShut  [1]{\csname bibitem#1\endcsname}%
\let\auto@bib@innerbib\@empty
\bibitem [{\citenamefont {Cartan}(1923)}]{cartan_1923}%
  \BibitemOpen
  \bibfield  {author} {\bibinfo {author} {\bibfnamefont {E.}~\bibnamefont
  {Cartan}},\ }\href@noop {} {\bibfield  {journal} {\bibinfo  {journal}
  {Annales Sci.Ecole Norm.Sup.}\ }\textbf {\bibinfo {volume} {40}},\ \bibinfo
  {pages} {325} (\bibinfo {year} {1923})}\BibitemShut {NoStop}%
\bibitem [{\citenamefont {Goudsmit}\ and\ \citenamefont
  {Uhlenbeck}()}]{goudsmit_1925}%
  \BibitemOpen
  \bibfield  {author} {\bibinfo {author} {\bibfnamefont {S.}~\bibnamefont
  {Goudsmit}}\ and\ \bibinfo {author} {\bibfnamefont {G.}~\bibnamefont
  {Uhlenbeck}},\ }\href@noop {} {\bibfield  {journal} {\bibinfo  {journal}
  {Physica}\ }\textbf {\bibinfo {volume} {6}},\ \bibinfo {pages}
  {273}}\BibitemShut {NoStop}%
\bibitem [{\citenamefont {Dirac}(1928)}]{dirac_1928}%
  \BibitemOpen
  \bibfield  {author} {\bibinfo {author} {\bibfnamefont {P.~A.~M.}\
  \bibnamefont {Dirac}},\ }\href {\doibase 10.1098/rspa.1928.0023} {\bibfield
  {journal} {\bibinfo  {journal} {Proceedings of the Royal Society of London A:
  Mathematical, Physical and Engineering Sciences}\ }\textbf {\bibinfo {volume}
  {117}},\ \bibinfo {pages} {610} (\bibinfo {year} {1928})}\BibitemShut
  {NoStop}%
\bibitem [{\citenamefont {Shankar}(1994)}]{shankar_1994}%
  \BibitemOpen
  \bibfield  {author} {\bibinfo {author} {\bibfnamefont {R.}~\bibnamefont
  {Shankar}},\ }\href@noop {} {\emph {\bibinfo {title} {Principles of Quantum
  Mechanics}}}\ (\bibinfo  {publisher} {Springer},\ \bibinfo {year}
  {1994})\BibitemShut {NoStop}%
\bibitem [{\citenamefont {Sakurai}(1994)}]{sakurai_1994}%
  \BibitemOpen
  \bibfield  {author} {\bibinfo {author} {\bibfnamefont {J.~J.}\ \bibnamefont
  {Sakurai}},\ }\href@noop {} {\emph {\bibinfo {title} {Modern Quantum
  Mechanics}}}\ (\bibinfo  {publisher} {Addison Wesley},\ \bibinfo {year}
  {1994})\BibitemShut {NoStop}%
\bibitem [{\citenamefont {Nyborg}(1962)}]{nyborg_1961A}%
  \BibitemOpen
  \bibfield  {author} {\bibinfo {author} {\bibfnamefont {P.}~\bibnamefont
  {Nyborg}},\ }\href {\doibase 10.1007/BF02733541} {\bibfield  {journal}
  {\bibinfo  {journal} {Il Nuovo Cimento Series 10}\ }\textbf {\bibinfo
  {volume} {23}},\ \bibinfo {pages} {47} (\bibinfo {year} {1962})}\BibitemShut
  {NoStop}%
\bibitem [{\citenamefont {{{\'S}redniawa}}(1996)}]{review_1996}%
  \BibitemOpen
  \bibfield  {author} {\bibinfo {author} {\bibfnamefont {B.}~\bibnamefont
  {{{\'S}redniawa}}},\ }\href@noop {} {\bibfield  {journal} {\bibinfo
  {journal} {Acta Cosmologica}\ }\textbf {\bibinfo {volume} {22}},\ \bibinfo
  {pages} {91} (\bibinfo {year} {1996})}\BibitemShut {NoStop}%
\bibitem [{\citenamefont {Frydryszak}(1996)}]{lagrangian_1996}%
  \BibitemOpen
  \bibfield  {author} {\bibinfo {author} {\bibfnamefont {A.}~\bibnamefont
  {Frydryszak}},\ }\href@noop {} {\  (\bibinfo {year} {1996})},\ \Eprint
  {http://arxiv.org/abs/hep-th/9601020} {arXiv:hep-th/9601020 [hep-th]}
  \BibitemShut {NoStop}%
\bibitem [{\citenamefont {Gaioli}\ and\ \citenamefont
  {Garcia~Alvarez}(1998)}]{review_1998}%
  \BibitemOpen
  \bibfield  {author} {\bibinfo {author} {\bibfnamefont {F.~H.}\ \bibnamefont
  {Gaioli}}\ and\ \bibinfo {author} {\bibfnamefont {E.~T.}\ \bibnamefont
  {Garcia~Alvarez}},\ }\href {\doibase 10.1023/A:1018834217984} {\bibfield
  {journal} {\bibinfo  {journal} {Found.Phys.}\ }\textbf {\bibinfo {volume}
  {28}},\ \bibinfo {pages} {1539} (\bibinfo {year} {1998})},\ \Eprint
  {http://arxiv.org/abs/hep-th/9807131} {arXiv:hep-th/9807131 [hep-th]}
  \BibitemShut {NoStop}%
\bibitem [{\citenamefont {Corben}(1968)}]{corbenbook_1968}%
  \BibitemOpen
  \bibfield  {author} {\bibinfo {author} {\bibfnamefont {H.}~\bibnamefont
  {Corben}},\ }\href@noop {} {\emph {\bibinfo {title} {{Classical and Quantum
  Theories of Spin}}}}\ (\bibinfo  {publisher} {Holden--Day},\ \bibinfo {year}
  {1968})\BibitemShut {NoStop}%
\bibitem [{\citenamefont {Rivas}(2002)}]{rivas_2002}%
  \BibitemOpen
  \bibfield  {author} {\bibinfo {author} {\bibfnamefont {M.}~\bibnamefont
  {Rivas}},\ }\href@noop {} {}Fundamental Theories of Physics (Book 116)\
  (\bibinfo {year} {12002})\BibitemShut {NoStop}%
\bibitem [{\citenamefont {Frenkel}(1926)}]{frenkel_1926}%
  \BibitemOpen
  \bibfield  {author} {\bibinfo {author} {\bibfnamefont {J.}~\bibnamefont
  {Frenkel}},\ }\href@noop {} {\bibfield  {journal} {\bibinfo  {journal} {Z.
  Physik}\ }\textbf {\bibinfo {volume} {37}},\ \bibinfo {pages} {243} (\bibinfo
  {year} {1926})}\BibitemShut {NoStop}%
\bibitem [{\citenamefont {Thomas}(1926)}]{thomas_1926}%
  \BibitemOpen
  \bibfield  {author} {\bibinfo {author} {\bibfnamefont {L.}~\bibnamefont
  {Thomas}},\ }\href {\doibase 10.1038/117514a0} {\bibfield  {journal}
  {\bibinfo  {journal} {Nature}\ }\textbf {\bibinfo {volume} {117}},\ \bibinfo
  {pages} {514} (\bibinfo {year} {1926})}\BibitemShut {NoStop}%
\bibitem [{\citenamefont {Thomas}(1927)}]{thomas_1927}%
  \BibitemOpen
  \bibfield  {author} {\bibinfo {author} {\bibfnamefont {L.}~\bibnamefont
  {Thomas}},\ }\href@noop {} {\bibfield  {journal} {\bibinfo  {journal}
  {Phil.Mag.}\ }\textbf {\bibinfo {volume} {3}},\ \bibinfo {pages} {1}
  (\bibinfo {year} {1927})}\BibitemShut {NoStop}%
\bibitem [{\citenamefont {Kramers}(1934)}]{kramers_1934}%
  \BibitemOpen
  \bibfield  {author} {\bibinfo {author} {\bibfnamefont {H.}~\bibnamefont
  {Kramers}},\ }\href {\doibase
  http://dx.doi.org/10.1016/S0031-8914(34)80276-5} {\bibfield  {journal}
  {\bibinfo  {journal} {Physica}\ }\textbf {\bibinfo {volume} {1}},\ \bibinfo
  {pages} {825 } (\bibinfo {year} {1934})}\BibitemShut {NoStop}%
\bibitem [{\citenamefont {Kramers}(2003)}]{kramers_QM}%
  \BibitemOpen
  \bibfield  {author} {\bibinfo {author} {\bibfnamefont {H.}~\bibnamefont
  {Kramers}},\ }\href@noop {} {\emph {\bibinfo {title} {{Quantum Mechanics}}}}\
  (\bibinfo  {publisher} {Dover Publications},\ \bibinfo {year} {2003})\ Chap.\
  \bibinfo {chapter} {The Spinning Electron}\BibitemShut {NoStop}%
\bibitem [{\citenamefont {Mathisson}(1937)}]{mathisson_1937}%
  \BibitemOpen
  \bibfield  {author} {\bibinfo {author} {\bibfnamefont {M.}~\bibnamefont
  {Mathisson}},\ }\href@noop {} {\bibfield  {journal} {\bibinfo  {journal}
  {Acta Phys.Polon.}\ }\textbf {\bibinfo {volume} {6}},\ \bibinfo {pages} {163}
  (\bibinfo {year} {1937})}\BibitemShut {NoStop}%
\bibitem [{\citenamefont {Papapetrou}(1951)}]{papapetrou_1951A}%
  \BibitemOpen
  \bibfield  {author} {\bibinfo {author} {\bibfnamefont {A.}~\bibnamefont
  {Papapetrou}},\ }\href {\doibase 10.1098/rspa.1951.0200} {\bibfield
  {journal} {\bibinfo  {journal} {Proc.Roy.Soc.Lond.}\ }\textbf {\bibinfo
  {volume} {A209}},\ \bibinfo {pages} {248} (\bibinfo {year}
  {1951})}\BibitemShut {NoStop}%
\bibitem [{\citenamefont {Corinaldesi}\ and\ \citenamefont
  {Papapetrou}(1951)}]{papapetrou_1951B}%
  \BibitemOpen
  \bibfield  {author} {\bibinfo {author} {\bibfnamefont {E.}~\bibnamefont
  {Corinaldesi}}\ and\ \bibinfo {author} {\bibfnamefont {A.}~\bibnamefont
  {Papapetrou}},\ }\href {\doibase 10.1098/rspa.1951.0201} {\bibfield
  {journal} {\bibinfo  {journal} {Proc.Roy.Soc.Lond.}\ }\textbf {\bibinfo
  {volume} {A209}},\ \bibinfo {pages} {259} (\bibinfo {year}
  {1951})}\BibitemShut {NoStop}%
\bibitem [{\citenamefont {Dixon}(1964)}]{dixon_1964}%
  \BibitemOpen
  \bibfield  {author} {\bibinfo {author} {\bibfnamefont {W.}~\bibnamefont
  {Dixon}},\ }\href {\doibase 10.1007/BF02734579} {\bibfield  {journal}
  {\bibinfo  {journal} {Il Nuovo Cimento}\ }\textbf {\bibinfo {volume} {34}},\
  \bibinfo {pages} {317} (\bibinfo {year} {1964})}\BibitemShut {NoStop}%
\bibitem [{\citenamefont {Dixon}(1967)}]{dixon_1967}%
  \BibitemOpen
  \bibfield  {author} {\bibinfo {author} {\bibfnamefont {W.~G.}\ \bibnamefont
  {Dixon}},\ }\href@noop {} {\bibfield  {journal} {\bibinfo  {journal} {Journal
  of Mathematical Physics}\ }\textbf {\bibinfo {volume} {8}} (\bibinfo {year}
  {1967})}\BibitemShut {NoStop}%
\bibitem [{\citenamefont {Dixon}(1970)}]{dixon_1970}%
  \BibitemOpen
  \bibfield  {author} {\bibinfo {author} {\bibfnamefont {W.~G.}\ \bibnamefont
  {Dixon}},\ }\href {\doibase 10.1098/rspa.1970.0020} {\bibfield  {journal}
  {\bibinfo  {journal} {Proceedings of the Royal Society of London A:
  Mathematical, Physical and Engineering Sciences}\ }\textbf {\bibinfo {volume}
  {314}},\ \bibinfo {pages} {499} (\bibinfo {year} {1970})}\BibitemShut
  {NoStop}%
\bibitem [{\citenamefont {Bhabha}\ and\ \citenamefont
  {Corben}(1941)}]{corben_1941}%
  \BibitemOpen
  \bibfield  {author} {\bibinfo {author} {\bibfnamefont {H.~J.}\ \bibnamefont
  {Bhabha}}\ and\ \bibinfo {author} {\bibfnamefont {H.~C.}\ \bibnamefont
  {Corben}},\ }\href {\doibase 10.1098/rspa.1941.0056} {\ \textbf {\bibinfo
  {volume} {178}},\ \bibinfo {pages} {273} (\bibinfo {year}
  {1941})}\BibitemShut {NoStop}%
\bibitem [{\citenamefont {Corben}(1961{\natexlab{a}})}]{corben_1960}%
  \BibitemOpen
  \bibfield  {author} {\bibinfo {author} {\bibfnamefont {H.}~\bibnamefont
  {Corben}},\ }\href {\doibase 10.1103/PhysRev.121.1833} {\bibfield  {journal}
  {\bibinfo  {journal} {Phys.Rev.}\ }\textbf {\bibinfo {volume} {121}},\
  \bibinfo {pages} {1833} (\bibinfo {year} {1961}{\natexlab{a}})}\BibitemShut
  {NoStop}%
\bibitem [{\citenamefont {Corben}(1961{\natexlab{b}})}]{corben_1961}%
  \BibitemOpen
  \bibfield  {author} {\bibinfo {author} {\bibfnamefont {H.}~\bibnamefont
  {Corben}},\ }\href {\doibase 10.1007/BF02731501} {\bibfield  {journal}
  {\bibinfo  {journal} {Il Nuovo Cimento Series 10}\ }\textbf {\bibinfo
  {volume} {20}},\ \bibinfo {pages} {529} (\bibinfo {year}
  {1961}{\natexlab{b}})}\BibitemShut {NoStop}%
\bibitem [{\citenamefont {Weyssenhoff}\ and\ \citenamefont
  {Raabe}(1947{\natexlab{a}})}]{weyss_1947A}%
  \BibitemOpen
  \bibfield  {author} {\bibinfo {author} {\bibfnamefont {J.}~\bibnamefont
  {Weyssenhoff}}\ and\ \bibinfo {author} {\bibfnamefont {A.}~\bibnamefont
  {Raabe}},\ }\href@noop {} {\bibfield  {journal} {\bibinfo  {journal} {Acta
  Phys. Pol.}\ }\textbf {\bibinfo {volume} {9}},\ \bibinfo {pages} {7}
  (\bibinfo {year} {1947}{\natexlab{a}})}\BibitemShut {NoStop}%
\bibitem [{\citenamefont {Weyssenhoff}\ and\ \citenamefont
  {Raabe}(1947{\natexlab{b}})}]{weyss_1947B}%
  \BibitemOpen
  \bibfield  {author} {\bibinfo {author} {\bibfnamefont {J.}~\bibnamefont
  {Weyssenhoff}}\ and\ \bibinfo {author} {\bibfnamefont {A.}~\bibnamefont
  {Raabe}},\ }\href@noop {} {\bibfield  {journal} {\bibinfo  {journal} {Acta
  Phys. Pol.}\ }\textbf {\bibinfo {volume} {9}},\ \bibinfo {pages} {19}
  (\bibinfo {year} {1947}{\natexlab{b}})}\BibitemShut {NoStop}%
\bibitem [{\citenamefont {{Barut}}(1958)}]{barut_1958}%
  \BibitemOpen
  \bibfield  {author} {\bibinfo {author} {\bibfnamefont {A.~O.}\ \bibnamefont
  {{Barut}}},\ }\href {\doibase 10.1016/0003-4916(58)90045-9} {\bibfield
  {journal} {\bibinfo  {journal} {Annals of Physics}\ }\textbf {\bibinfo
  {volume} {5}},\ \bibinfo {pages} {95} (\bibinfo {year} {1958})}\BibitemShut
  {NoStop}%
\bibitem [{\citenamefont {Barut}\ and\ \citenamefont
  {Duru}(1984)}]{barut_1984A}%
  \BibitemOpen
  \bibfield  {author} {\bibinfo {author} {\bibfnamefont {A.}~\bibnamefont
  {Barut}}\ and\ \bibinfo {author} {\bibfnamefont {I.}~\bibnamefont {Duru}},\
  }\href {\doibase 10.1103/PhysRevLett.53.2355} {\bibfield  {journal} {\bibinfo
   {journal} {Phys.Rev.Lett.}\ }\textbf {\bibinfo {volume} {53}},\ \bibinfo
  {pages} {2355} (\bibinfo {year} {1984})}\BibitemShut {NoStop}%
\bibitem [{\citenamefont {Barut}\ and\ \citenamefont
  {Zanghi}(1984)}]{barut_1984B}%
  \BibitemOpen
  \bibfield  {author} {\bibinfo {author} {\bibfnamefont {A.}~\bibnamefont
  {Barut}}\ and\ \bibinfo {author} {\bibfnamefont {N.}~\bibnamefont {Zanghi}},\
  }\href {\doibase 10.1103/PhysRevLett.52.2009} {\bibfield  {journal} {\bibinfo
   {journal} {Phys. Rev. Lett.}\ }\textbf {\bibinfo {volume} {52}},\ \bibinfo
  {pages} {2009} (\bibinfo {year} {1984})}\BibitemShut {NoStop}%
\bibitem [{\citenamefont {Barut}\ and\ \citenamefont
  {Duru}(1989)}]{barut_1988}%
  \BibitemOpen
  \bibfield  {author} {\bibinfo {author} {\bibfnamefont {A.}~\bibnamefont
  {Barut}}\ and\ \bibinfo {author} {\bibfnamefont {I.}~\bibnamefont {Duru}},\
  }\href {\doibase http://dx.doi.org/10.1016/0370-1573(89)90146-4} {\bibfield
  {journal} {\bibinfo  {journal} {Physics Reports}\ }\textbf {\bibinfo {volume}
  {172}},\ \bibinfo {pages} {1 } (\bibinfo {year} {1989})}\BibitemShut
  {NoStop}%
\bibitem [{\citenamefont {Deriglazov}(2013)}]{deriglazov_2012}%
  \BibitemOpen
  \bibfield  {author} {\bibinfo {author} {\bibfnamefont {A.}~\bibnamefont
  {Deriglazov}},\ }\href {\doibase 10.1142/S0217732312502343} {\bibfield
  {journal} {\bibinfo  {journal} {Mod.Phys.Lett.}\ }\textbf {\bibinfo {volume}
  {A28}},\ \bibinfo {pages} {1250234} (\bibinfo {year} {2013})},\ \Eprint
  {http://arxiv.org/abs/1204.2494} {arXiv:1204.2494 [hep-th]} \BibitemShut
  {NoStop}%
\bibitem [{\citenamefont {Deriglazov}(2014)}]{deriglazov_2014}%
  \BibitemOpen
  \bibfield  {author} {\bibinfo {author} {\bibfnamefont {A.~A.}\ \bibnamefont
  {Deriglazov}},\ }\href {\doibase 10.1016/j.physletb.2014.07.029} {\bibfield
  {journal} {\bibinfo  {journal} {Phys.Lett.}\ }\textbf {\bibinfo {volume}
  {B736}},\ \bibinfo {pages} {278} (\bibinfo {year} {2014})},\ \Eprint
  {http://arxiv.org/abs/1406.6715} {arXiv:1406.6715 [physics.gen-ph]}
  \BibitemShut {NoStop}%
\bibitem [{\citenamefont {Costa}\ \emph {et~al.}(2012)\citenamefont {Costa},
  \citenamefont {Herdeiro}, \citenamefont {Nat\'ario},\ and\ \citenamefont
  {Zilh\~ao}}]{costa_2012}%
  \BibitemOpen
  \bibfield  {author} {\bibinfo {author} {\bibfnamefont {L.}~\bibnamefont
  {Costa}}, \bibinfo {author} {\bibfnamefont {C.}~\bibnamefont {Herdeiro}},
  \bibinfo {author} {\bibfnamefont {J.}~\bibnamefont {Nat\'ario}}, \ and\
  \bibinfo {author} {\bibfnamefont {M.}~\bibnamefont {Zilh\~ao}},\ }\href
  {\doibase 10.1103/PhysRevD.85.024001} {\bibfield  {journal} {\bibinfo
  {journal} {Phys. Rev. D}\ }\textbf {\bibinfo {volume} {85}},\ \bibinfo
  {pages} {024001} (\bibinfo {year} {2012})}\BibitemShut {NoStop}%
\bibitem [{\citenamefont {Costa}\ and\ \citenamefont
  {Nat{\'a}rio}(2014)}]{costa_2014}%
  \BibitemOpen
  \bibfield  {author} {\bibinfo {author} {\bibfnamefont {L.~F.}\ \bibnamefont
  {Costa}}\ and\ \bibinfo {author} {\bibfnamefont {J.}~\bibnamefont
  {Nat{\'a}rio}},\ }\href@noop {} {\  (\bibinfo {year} {2014})},\ \Eprint
  {http://arxiv.org/abs/1410.6443} {arXiv:1410.6443 [gr-qc]} \BibitemShut
  {NoStop}%
\bibitem [{\citenamefont {Hanson}\ and\ \citenamefont
  {Regge}(1974)}]{hanson_1974}%
  \BibitemOpen
  \bibfield  {author} {\bibinfo {author} {\bibfnamefont {A.~J.}\ \bibnamefont
  {Hanson}}\ and\ \bibinfo {author} {\bibfnamefont {T.}~\bibnamefont {Regge}},\
  }\href {\doibase 10.1016/0003-4916(74)90046-3} {\bibfield  {journal}
  {\bibinfo  {journal} {Annals Phys.}\ }\textbf {\bibinfo {volume} {87}},\
  \bibinfo {pages} {498} (\bibinfo {year} {1974})}\BibitemShut {NoStop}%
\bibitem [{\citenamefont {Balachandran}\ \emph {et~al.}(1977)\citenamefont
  {Balachandran}, \citenamefont {Salomonson}, \citenamefont {Skagerstam},\ and\
  \citenamefont {Winnberg}}]{balachandran_1976}%
  \BibitemOpen
  \bibfield  {author} {\bibinfo {author} {\bibfnamefont {A.}~\bibnamefont
  {Balachandran}}, \bibinfo {author} {\bibfnamefont {P.}~\bibnamefont
  {Salomonson}}, \bibinfo {author} {\bibfnamefont {B.-S.}\ \bibnamefont
  {Skagerstam}}, \ and\ \bibinfo {author} {\bibfnamefont {J.-O.}\ \bibnamefont
  {Winnberg}},\ }\href {\doibase 10.1103/PhysRevD.15.2308} {\bibfield
  {journal} {\bibinfo  {journal} {Phys.Rev.}\ }\textbf {\bibinfo {volume}
  {D15}},\ \bibinfo {pages} {2308} (\bibinfo {year} {1977})}\BibitemShut
  {NoStop}%
\bibitem [{\citenamefont {Balachandran}\ \emph {et~al.}(1980)\citenamefont
  {Balachandran}, \citenamefont {Marmo}, \citenamefont {Skagerstam},\ and\
  \citenamefont {Stern}}]{balachandran_1979}%
  \BibitemOpen
  \bibfield  {author} {\bibinfo {author} {\bibfnamefont {A.}~\bibnamefont
  {Balachandran}}, \bibinfo {author} {\bibfnamefont {G.}~\bibnamefont {Marmo}},
  \bibinfo {author} {\bibfnamefont {B.}~\bibnamefont {Skagerstam}}, \ and\
  \bibinfo {author} {\bibfnamefont {A.}~\bibnamefont {Stern}},\ }\href
  {\doibase 10.1016/0370-2693(80)90009-X} {\bibfield  {journal} {\bibinfo
  {journal} {Phys.Lett.}\ }\textbf {\bibinfo {volume} {B89}},\ \bibinfo {pages}
  {199} (\bibinfo {year} {1980})}\BibitemShut {NoStop}%
\bibitem [{\citenamefont {Balachandran}\ \emph {et~al.}(1982)\citenamefont
  {Balachandran}, \citenamefont {Marmo}, \citenamefont {Mukunda}, \citenamefont
  {Nilsson}, \citenamefont {Simoni} \emph {et~al.}}]{balachandran_1981}%
  \BibitemOpen
  \bibfield  {author} {\bibinfo {author} {\bibfnamefont {A.}~\bibnamefont
  {Balachandran}}, \bibinfo {author} {\bibfnamefont {G.}~\bibnamefont {Marmo}},
  \bibinfo {author} {\bibfnamefont {N.}~\bibnamefont {Mukunda}}, \bibinfo
  {author} {\bibfnamefont {J.}~\bibnamefont {Nilsson}}, \bibinfo {author}
  {\bibfnamefont {A.}~\bibnamefont {Simoni}},  \emph {et~al.},\ }\href
  {\doibase 10.1007/BF02816669} {\bibfield  {journal} {\bibinfo  {journal}
  {Nuovo Cim.}\ }\textbf {\bibinfo {volume} {A67}},\ \bibinfo {pages} {121}
  (\bibinfo {year} {1982})}\BibitemShut {NoStop}%
\bibitem [{\citenamefont {Kirillov}(1976)}]{kirillov_1976}%
  \BibitemOpen
  \bibfield  {author} {\bibinfo {author} {\bibfnamefont {A.}~\bibnamefont
  {Kirillov}},\ }\href@noop {} {\emph {\bibinfo {title} {{Elements of the
  Theory of Representations}}}}\ (\bibinfo  {publisher} {Springer--Verlag.},\
  \bibinfo {year} {1976})\BibitemShut {NoStop}%
\bibitem [{\citenamefont {Kostant}(1970)}]{kostant_1970}%
  \BibitemOpen
  \bibfield  {author} {\bibinfo {author} {\bibfnamefont {B.}~\bibnamefont
  {Kostant}},\ }in\ \href {\doibase 10.1007/BFb0079068} {\emph {\bibinfo
  {booktitle} {Lectures in Modern Analysis and Applications III}}},\ \bibinfo
  {series} {Lecture Notes in Mathematics}, Vol.\ \bibinfo {volume} {170},\
  \bibinfo {editor} {edited by\ \bibinfo {editor} {\bibfnamefont
  {C.}~\bibnamefont {Taam}}}\ (\bibinfo  {publisher} {Springer Berlin
  Heidelberg},\ \bibinfo {year} {1970})\ pp.\ \bibinfo {pages}
  {87--208}\BibitemShut {NoStop}%
\bibitem [{\citenamefont {Souriau}(1970)}]{souriau_1970}%
  \BibitemOpen
  \bibfield  {author} {\bibinfo {author} {\bibfnamefont {J.}~\bibnamefont
  {Souriau}},\ }\href@noop {} {\emph {\bibinfo {title} {{Structure des Systemes
  Dynamiques}}}}\ (\bibinfo  {publisher} {Dunod},\ \bibinfo {year}
  {1970})\BibitemShut {NoStop}%
\bibitem [{\citenamefont {Souriau}(1997)}]{souriau_1997}%
  \BibitemOpen
  \bibfield  {author} {\bibinfo {author} {\bibfnamefont {J.}~\bibnamefont
  {Souriau}},\ }\href@noop {} {\emph {\bibinfo {title} {{Structure of Dynamical
  Systems: A symplectic view of physics}}}}\ (\bibinfo  {publisher} {Birkhauser
  Boston.},\ \bibinfo {year} {1997})\BibitemShut {NoStop}%
\bibitem [{\citenamefont {Nielsen}\ and\ \citenamefont
  {Rohrlich}(1988)}]{nielsen_1987}%
  \BibitemOpen
  \bibfield  {author} {\bibinfo {author} {\bibfnamefont {H.}~\bibnamefont
  {Nielsen}}\ and\ \bibinfo {author} {\bibfnamefont {D.}~\bibnamefont
  {Rohrlich}},\ }\href {\doibase
  http://dx.doi.org/10.1016/0550-3213(88)90545-7} {\bibfield  {journal}
  {\bibinfo  {journal} {Nuclear Physics B}\ }\textbf {\bibinfo {volume}
  {299}},\ \bibinfo {pages} {471 } (\bibinfo {year} {1988})}\BibitemShut
  {NoStop}%
\bibitem [{\citenamefont {Alekseev}\ \emph {et~al.}(1988)\citenamefont
  {Alekseev}, \citenamefont {Faddeev},\ and\ \citenamefont
  {Shatashvili}}]{faddeev_1988}%
  \BibitemOpen
  \bibfield  {author} {\bibinfo {author} {\bibfnamefont {A.}~\bibnamefont
  {Alekseev}}, \bibinfo {author} {\bibfnamefont {L.}~\bibnamefont {Faddeev}}, \
  and\ \bibinfo {author} {\bibfnamefont {S.}~\bibnamefont {Shatashvili}},\
  }\href {\doibase http://dx.doi.org/10.1016/0393-0440(88)90031-9} {\bibfield
  {journal} {\bibinfo  {journal} {Journal of Geometry and Physics}\ }\textbf
  {\bibinfo {volume} {5}},\ \bibinfo {pages} {391 } (\bibinfo {year}
  {1988})}\BibitemShut {NoStop}%
\bibitem [{\citenamefont {Johnson}(1989)}]{johnson_1989}%
  \BibitemOpen
  \bibfield  {author} {\bibinfo {author} {\bibfnamefont {K.}~\bibnamefont
  {Johnson}},\ }\href {\doibase http://dx.doi.org/10.1016/0003-4916(89)90120-6}
  {\bibfield  {journal} {\bibinfo  {journal} {Annals of Physics}\ }\textbf
  {\bibinfo {volume} {192}},\ \bibinfo {pages} {104 } (\bibinfo {year}
  {1989})}\BibitemShut {NoStop}%
\bibitem [{\citenamefont {Wiegmann}(1989)}]{wiegmann_1989B}%
  \BibitemOpen
  \bibfield  {author} {\bibinfo {author} {\bibfnamefont {P.}~\bibnamefont
  {Wiegmann}},\ }\href {\doibase 10.1016/0550-3213(89)90144-2} {\bibfield
  {journal} {\bibinfo  {journal} {Nucl.Phys.}\ }\textbf {\bibinfo {volume}
  {B323}},\ \bibinfo {pages} {311} (\bibinfo {year} {1989})}\BibitemShut
  {NoStop}%
\bibitem [{\citenamefont {Mauro}(2004)}]{mauro_2004}%
  \BibitemOpen
  \bibfield  {author} {\bibinfo {author} {\bibfnamefont {D.}~\bibnamefont
  {Mauro}},\ }\href {\doibase http://dx.doi.org/10.1016/j.physletb.2004.07.016}
  {\bibfield  {journal} {\bibinfo  {journal} {Physics Letters B}\ }\textbf
  {\bibinfo {volume} {597}},\ \bibinfo {pages} {94 } (\bibinfo {year}
  {2004})}\BibitemShut {NoStop}%
\bibitem [{\citenamefont {Feynman}(1950)}]{feynman_1950}%
  \BibitemOpen
  \bibfield  {author} {\bibinfo {author} {\bibfnamefont {R.~P.}\ \bibnamefont
  {Feynman}},\ }\href {\doibase 10.1103/PhysRev.80.440} {\bibfield  {journal}
  {\bibinfo  {journal} {Phys. Rev.}\ }\textbf {\bibinfo {volume} {80}},\
  \bibinfo {pages} {440} (\bibinfo {year} {1950})}\BibitemShut {NoStop}%
\bibitem [{\citenamefont {Feynman}(1951)}]{feynman_1951}%
  \BibitemOpen
  \bibfield  {author} {\bibinfo {author} {\bibfnamefont {R.~P.}\ \bibnamefont
  {Feynman}},\ }\href {\doibase 10.1103/PhysRev.84.108} {\bibfield  {journal}
  {\bibinfo  {journal} {Phys. Rev.}\ }\textbf {\bibinfo {volume} {84}},\
  \bibinfo {pages} {108} (\bibinfo {year} {1951})}\BibitemShut {NoStop}%
\bibitem [{\citenamefont {Fradkin}(1966)}]{fradkin_1965}%
  \BibitemOpen
  \bibfield  {author} {\bibinfo {author} {\bibfnamefont {E.}~\bibnamefont
  {Fradkin}},\ }\href {\doibase http://dx.doi.org/10.1016/0029-5582(66)90200-8}
  {\bibfield  {journal} {\bibinfo  {journal} {Nuclear Physics}\ }\textbf
  {\bibinfo {volume} {76}},\ \bibinfo {pages} {588 } (\bibinfo {year}
  {1966})}\BibitemShut {NoStop}%
\bibitem [{\citenamefont {Berezin}\ and\ \citenamefont
  {Marinov}(1977)}]{berezin_1977}%
  \BibitemOpen
  \bibfield  {author} {\bibinfo {author} {\bibfnamefont {F.}~\bibnamefont
  {Berezin}}\ and\ \bibinfo {author} {\bibfnamefont {M.}~\bibnamefont
  {Marinov}},\ }\href {\doibase http://dx.doi.org/10.1016/0003-4916(77)90335-9}
  {\bibfield  {journal} {\bibinfo  {journal} {Annals of Physics}\ }\textbf
  {\bibinfo {volume} {104}},\ \bibinfo {pages} {336 } (\bibinfo {year}
  {1977})}\BibitemShut {NoStop}%
\bibitem [{\citenamefont {Howe}\ \emph {et~al.}(1988)\citenamefont {Howe},
  \citenamefont {Penati}, \citenamefont {Pernici},\ and\ \citenamefont
  {Townsend}}]{howe_1976}%
  \BibitemOpen
  \bibfield  {author} {\bibinfo {author} {\bibfnamefont {P.}~\bibnamefont
  {Howe}}, \bibinfo {author} {\bibfnamefont {S.}~\bibnamefont {Penati}},
  \bibinfo {author} {\bibfnamefont {M.}~\bibnamefont {Pernici}}, \ and\
  \bibinfo {author} {\bibfnamefont {P.}~\bibnamefont {Townsend}},\ }\href
  {\doibase http://dx.doi.org/10.1016/0370-2693(88)91358-5} {\bibfield
  {journal} {\bibinfo  {journal} {Physics Letters B}\ }\textbf {\bibinfo
  {volume} {215}},\ \bibinfo {pages} {555 } (\bibinfo {year}
  {1988})}\BibitemShut {NoStop}%
\bibitem [{\citenamefont {Brink}\ \emph {et~al.}(1977)\citenamefont {Brink},
  \citenamefont {Vecchia},\ and\ \citenamefont {Howe}}]{brink_1976}%
  \BibitemOpen
  \bibfield  {author} {\bibinfo {author} {\bibfnamefont {L.}~\bibnamefont
  {Brink}}, \bibinfo {author} {\bibfnamefont {P.~D.}\ \bibnamefont {Vecchia}},
  \ and\ \bibinfo {author} {\bibfnamefont {P.}~\bibnamefont {Howe}},\ }\href
  {\doibase http://dx.doi.org/10.1016/0550-3213(77)90364-9} {\bibfield
  {journal} {\bibinfo  {journal} {Nuclear Physics B}\ }\textbf {\bibinfo
  {volume} {118}},\ \bibinfo {pages} {76 } (\bibinfo {year}
  {1977})}\BibitemShut {NoStop}%
\bibitem [{\citenamefont {Schubert}(2001)}]{schubert_2000}%
  \BibitemOpen
  \bibfield  {author} {\bibinfo {author} {\bibfnamefont {C.}~\bibnamefont
  {Schubert}},\ }\href {\doibase 10.1063/1.1374964} {\bibfield  {journal}
  {\bibinfo  {journal} {AIP Conf.Proc.}\ }\textbf {\bibinfo {volume} {564}},\
  \bibinfo {pages} {28} (\bibinfo {year} {2001})},\ \Eprint
  {http://arxiv.org/abs/hep-ph/0011331} {arXiv:hep-ph/0011331 [hep-ph]}
  \BibitemShut {NoStop}%
\bibitem [{\citenamefont {Amelino-Camelia}\ \emph
  {et~al.}(2011{\natexlab{a}})\citenamefont {Amelino-Camelia}, \citenamefont
  {Freidel}, \citenamefont {Kowalski-Glikman},\ and\ \citenamefont
  {Smolin}}]{relative_locality}%
  \BibitemOpen
  \bibfield  {author} {\bibinfo {author} {\bibfnamefont {G.}~\bibnamefont
  {Amelino-Camelia}}, \bibinfo {author} {\bibfnamefont {L.}~\bibnamefont
  {Freidel}}, \bibinfo {author} {\bibfnamefont {J.}~\bibnamefont
  {Kowalski-Glikman}}, \ and\ \bibinfo {author} {\bibfnamefont
  {L.}~\bibnamefont {Smolin}},\ }\href {\doibase 10.1103/PhysRevD.84.084010}
  {\bibfield  {journal} {\bibinfo  {journal} {Phys.Rev.}\ }\textbf {\bibinfo
  {volume} {D84}},\ \bibinfo {pages} {084010} (\bibinfo {year}
  {2011}{\natexlab{a}})},\ \Eprint {http://arxiv.org/abs/1101.0931}
  {arXiv:1101.0931 [hep-th]} \BibitemShut {NoStop}%
\bibitem [{\citenamefont {Freidel}\ and\ \citenamefont
  {Rempel}(2013{\natexlab{a}})}]{action_vertices}%
  \BibitemOpen
  \bibfield  {author} {\bibinfo {author} {\bibfnamefont {L.}~\bibnamefont
  {Freidel}}\ and\ \bibinfo {author} {\bibfnamefont {T.}~\bibnamefont
  {Rempel}},\ }\href@noop {} {\  (\bibinfo {year} {2013}{\natexlab{a}})},\
  \Eprint {http://arxiv.org/abs/1312.5396} {arXiv:1312.5396 [hep-th]}
  \BibitemShut {NoStop}%
\bibitem [{\citenamefont {Amelino-Camelia}\ \emph
  {et~al.}(2011{\natexlab{b}})\citenamefont {Amelino-Camelia}, \citenamefont
  {Freidel}, \citenamefont {Kowalski-Glikman},\ and\ \citenamefont
  {Smolin}}]{deepening}%
  \BibitemOpen
  \bibfield  {author} {\bibinfo {author} {\bibfnamefont {G.}~\bibnamefont
  {Amelino-Camelia}}, \bibinfo {author} {\bibfnamefont {L.}~\bibnamefont
  {Freidel}}, \bibinfo {author} {\bibfnamefont {J.}~\bibnamefont
  {Kowalski-Glikman}}, \ and\ \bibinfo {author} {\bibfnamefont
  {L.}~\bibnamefont {Smolin}},\ }\href {\doibase 10.1142/S0218271811020743,
  10.1007/s10714-011-1212-8} {\bibfield  {journal} {\bibinfo  {journal}
  {Gen.Rel.Grav.}\ }\textbf {\bibinfo {volume} {43}},\ \bibinfo {pages} {2547}
  (\bibinfo {year} {2011}{\natexlab{b}})},\ \Eprint
  {http://arxiv.org/abs/1106.0313} {arXiv:1106.0313 [hep-th]} \BibitemShut
  {NoStop}%
\bibitem [{\citenamefont {Freidel}\ and\ \citenamefont
  {Rempel}(2013{\natexlab{b}})}]{scalar_curved}%
  \BibitemOpen
  \bibfield  {author} {\bibinfo {author} {\bibfnamefont {L.}~\bibnamefont
  {Freidel}}\ and\ \bibinfo {author} {\bibfnamefont {T.}~\bibnamefont
  {Rempel}},\ }\href@noop {} {\  (\bibinfo {year} {2013}{\natexlab{b}})},\
  \Eprint {http://arxiv.org/abs/1312.3674} {arXiv:1312.3674 [hep-th]}
  \BibitemShut {NoStop}%
\bibitem [{\citenamefont {Freidel}\ \emph {et~al.}(2007)\citenamefont
  {Freidel}, \citenamefont {Girelli},\ and\ \citenamefont
  {Livine}}]{Freidel:2007qk}%
  \BibitemOpen
  \bibfield  {author} {\bibinfo {author} {\bibfnamefont {L.}~\bibnamefont
  {Freidel}}, \bibinfo {author} {\bibfnamefont {F.}~\bibnamefont {Girelli}}, \
  and\ \bibinfo {author} {\bibfnamefont {E.~R.}\ \bibnamefont {Livine}},\
  }\href {\doibase 10.1103/PhysRevD.75.105016} {\bibfield  {journal} {\bibinfo
  {journal} {Phys. Rev.}\ }\textbf {\bibinfo {volume} {D75}},\ \bibinfo {pages}
  {105016} (\bibinfo {year} {2007})},\ \Eprint
  {http://arxiv.org/abs/hep-th/0701113} {arXiv:hep-th/0701113 [hep-th]}
  \BibitemShut {NoStop}%
\bibitem [{\citenamefont {Kirillov}(2004)}]{kirillov_2002}%
  \BibitemOpen
  \bibfield  {author} {\bibinfo {author} {\bibfnamefont {A.}~\bibnamefont
  {Kirillov}},\ }\href@noop {} {\emph {\bibinfo {title} {{Lectures on the Orbit
  Method}}}}\ (\bibinfo  {publisher} {American Mathematical Society.},\
  \bibinfo {year} {2004})\BibitemShut {NoStop}%
\bibitem [{\citenamefont {Woodhouse}(1991)}]{woodhouse_1991}%
  \BibitemOpen
  \bibfield  {author} {\bibinfo {author} {\bibfnamefont {N.}~\bibnamefont
  {Woodhouse}},\ }\href@noop {} {\emph {\bibinfo {title} {{Geometric
  Quantization}}}}\ (\bibinfo  {publisher} {Oxford University Press.},\
  \bibinfo {year} {1991})\BibitemShut {NoStop}%
\bibitem [{\citenamefont {Bargmann}\ and\ \citenamefont
  {Wigner}(1948)}]{wigner_1948}%
  \BibitemOpen
  \bibfield  {author} {\bibinfo {author} {\bibfnamefont {V.}~\bibnamefont
  {Bargmann}}\ and\ \bibinfo {author} {\bibfnamefont {E.~P.}\ \bibnamefont
  {Wigner}},\ }\href {\doibase 10.1073/pnas.34.5.211} {\bibfield  {journal}
  {\bibinfo  {journal} {Proc.Nat.Acad.Sci.}\ }\textbf {\bibinfo {volume}
  {34}},\ \bibinfo {pages} {211} (\bibinfo {year} {1948})}\BibitemShut
  {NoStop}%
\bibitem [{\citenamefont {Edgren}\ \emph {et~al.}(2005)\citenamefont {Edgren},
  \citenamefont {Marnelius},\ and\ \citenamefont {Salomonson}}]{edgren_2005}%
  \BibitemOpen
  \bibfield  {author} {\bibinfo {author} {\bibfnamefont {L.}~\bibnamefont
  {Edgren}}, \bibinfo {author} {\bibfnamefont {R.}~\bibnamefont {Marnelius}}, \
  and\ \bibinfo {author} {\bibfnamefont {P.}~\bibnamefont {Salomonson}},\
  }\href {\doibase 10.1088/1126-6708/2005/05/002} {\bibfield  {journal}
  {\bibinfo  {journal} {JHEP}\ }\textbf {\bibinfo {volume} {0505}},\ \bibinfo
  {pages} {002} (\bibinfo {year} {2005})},\ \Eprint
  {http://arxiv.org/abs/hep-th/0503136} {arXiv:hep-th/0503136 [hep-th]}
  \BibitemShut {NoStop}%
\end{thebibliography}%

\end{document}